\documentclass[12pt]{article}
\usepackage{amsmath}
\usepackage{graphicx}
\usepackage{enumerate}
\usepackage{natbib}
\usepackage{url} 
\usepackage{lineno}
\usepackage{amssymb}
\usepackage{amsthm}
\usepackage{authblk}
\usepackage{epsfig}
\usepackage{graphics}
\usepackage{float}
\usepackage{subfigure}
\usepackage{multirow}
\usepackage{xcolor}
\usepackage{makeidx}
\usepackage{xspace}
\usepackage{wrapfig}
\usepackage{blindtext}
\usepackage{setspace}
\usepackage{algorithm}
\usepackage{algorithmic}
\usepackage[colorlinks=true,citecolor=blue,linkcolor=blue,urlcolor=blue]{hyperref}
\usepackage{booktabs}
\usepackage{bm}
\makeindex

\renewcommand{\algorithmicrequire}{ \textbf{Input:}} 
\renewcommand{\algorithmicensure}{ \textbf{Output:}} 

\definecolor{darkred}{rgb}{1, 0.1, 0.3}
\definecolor{darkblue}{rgb}{0.1, 0.1, 1}
\definecolor{darkgreen}{rgb}{0,0.6,0.5}

\newcommand{\blind}{1}

\begin{document}

\def\spacingset#1{\renewcommand{\baselinestretch}%
{#1}\small\normalsize} \spacingset{1}


\if1\blind
{
  \title{\bf Online stochastic generators using Slepian bases for regional bivariate wind speed ensembles from ERA5}
  \author{Yan Song\thanks{
    Corresponding author; email: yan.song@kaust.edu.sa}\hspace{.2cm}\\
    \vspace{-10pt}
    Statistics Program, King Abdullah University of Science and Technology, Saudi Arabia\\
    \vspace{5pt}
     Zubair Khalid \\
    School of Science and Engineering, Lahore University of Management Sciences, Pakistan\\
    \vspace{5pt}
    Marc G. Genton \\
     Statistics Program, King Abdullah University of Science and Technology, Saudi Arabia}
  \maketitle
} \fi

\if0\blind
{
\title{\bf Online stochastic generators using Slepian bases for regional bivariate wind speed ensembles from ERA5}
  \author{ 
    }
  \maketitle
}\fi

\vspace{-20pt}
\begin{abstract}
Reanalysis data, such as ERA5, provide a comprehensive and detailed representation of the Earth's system by assimilating observations into climate models. While crucial for climate research, they pose significant challenges in terms of generation, storage, and management. For 3-hourly bivariate wind speed ensembles from ERA5, which face these challenges, this paper proposes an online stochastic generator (OSG) applicable to any global region, offering fast stochastic approximations while storing only model parameters. A key innovation is the incorporation of the online updating, which allows data to sequentially enter the model in blocks of time and contribute to parameter updates. This approach reduces storage demands during modeling by eliminating the need to store and analyze the entire dataset, and enables near real-time emulations that complement the generation of reanalysis data. The Slepian concentration technique supports the efficiency of the proposed OSG by representing the data in a lower-dimensional space spanned by data-independent Slepian bases optimally concentrated within the specified region. We demonstrate the flexibility and efficiency of the OSG through two case studies requiring long and short blocks, specified for the Arabian-Peninsula region (ARP). For both cases, the OSG performs well across several statistical metrics and is comparable to the SG trained on the full dataset.
\end{abstract}

\noindent%


\newpage
\spacingset{1.8} 
\section{Introduction}
\label{sec:Introduction}
A reanalysis assimilates observations from diverse sources, such as satellites, weather stations, and buoys, into numerical weather prediction (NWP) models that simulate the physical dynamics of the atmosphere, oceans, and land surface. This process generates high-resolution reanalysis data spanning the past several decades and encompassing hundreds of climate variables. By combining real-world observations with fundamental physical laws, reanalysis data offer a comprehensive, consistent, and detailed description of the Earth's system. Reanalysis data have found wide applications, including climate research \citep{trenberth2011atmospheric}, weather forecasting \citep{cornejo2017wind} and environmental monitoring \citep{matsueda2015global}. 

Recent global reanalysis projects include the Climate Forecast System Version 2 \citep[CFSR,][]{saha2014ncep}, the Japanese 55-Year Reanalysis \citep[JRA-55,][]{kobayashi2015jra}, the Modern-Era Retrospective Analysis for Research and Applications Version 2 \citep[MERRA-2,][]{gelaro2017modern}, and the fifth generation European Centre for Medium-Range Weather Forecasts (ECMWF) Reanalysis \citep[ERA5,][]{ERA52020}, with the last two projects providing near real-time data. These reanalysis data are publicly accessible through online portals and application programming interfaces (APIs), enabling users to subset the data by variable, region and time period. Additionally, regional projects like the North American Regional Reanalysis \citep[NARR,][]{mesinger2006north} and the Uncertainties in Ensembles of Regional Reanalyses \citep[UERRA,][]{unden2016uerra} contribute insights into localized climate trend, variability, and extreme events. 

Nevertheless, reanalysis data encounter several limitations. Firstly, the reliability of reanalysis data depends on the collection and integration of observations, which can vary across regions, periods, and variables. Multiple ensembles of reanalysis data help assess the uncertainty. For instance, the uncertainty for ERA5 refers to a 10-member Ensemble of Data Assimilations (EDA) system. Secondly, the generation of reanalysis data, which involves integrating big volume observations from various sources and running NWP models, requires the support of high-performance computing, such as the Cray XC40 supercomputer used by ECMWF. Moreover, storing and managing high-resolution, petabyte-scale reanalysis data, especially those that are continuously updated in real-time, necessitates advanced hardware, sophisticated software, and efficient data management practices. For example, due to cost and computational considerations, the EDA for ERA5 operates at a lower spatial and temporal resolution (approximately $0.5$ degree horizontally and 3-hourly) compared to ERA5 itself (approximately $0.25$ degree horizontally and hourly).

Stochastic generators \citep[SGs,][]{jeong2018reducing} offer a solution to the aforementioned limitations. As a type of \textit{emulator} \citep{Sacks1989,Kennedy2001,Castruccio2013,Hu2021Approximating}, SGs are statistical models trained using several ensembles of existing data and serve as practical surrogates for data generation mechanisms, producing rapid stochastic approximations of the training data, referred to as \textit{emulations}. By storing only the model parameters rather than the entire dataset, SGs significantly reduce storage requirements. The idea of SGs has been explored in several studies. \citet{Castruccio2013}, \citet{castruccio2014statistical}, \citet{castruccio2016compressing}, and \citet{castruccio2017evolutionary} developed SGs to analyze the internal variability in annual temperature simulations generated by various Earth system models (ESMs). \citet{jeong2018reducing} introduced the formal concept of SGs for the first time to generate emulations for annual wind speed simulations. \citet{jeong2019stochastic} incorporated the Tukey g-and-h (TGH) transformation into an SG to handle the non-Gaussian characteristics of monthly wind speed simulations. \citet{Tagle2020} proposed a regional SG to investigate wind energy in Saudi Arabia with kilometer-scale spatial resolution and daily temporal resolution. \citet{Huang'sEmulator} provided an overview of SG techniques across various spatial and temporal scales, particularly for temperature simulations from the Community Earth System Model Large Ensemble \citep[CESM-LENS,][]{kay2015community}. Furthermore, leveraging the spherical harmonic transformation (SHT), \citet{song2024efficient} proposed an efficient SG capable of emulating even daily temperature simulations from the newly published CESM version 2 Large Ensembles (CESM2-LENS2) and validated its exceptional performance through various metrics.

The majority of existing SGs are designed for global climate simulations (i.e., on a sphere) of a single variable over a fixed period, with a spatial resolution of up to around $1$ degree horizontally and a temporal resolution of up to the daily scale. However, they are inadequate for emulating reanalysis data, which could be regional, multivariate, continuously arriving with higher spatio-temporal resolutions, and hence storage-consuming. For reanalysis data, a desirable SG should meet two key requirements. First, given storage limitations, it should efficiently model and emulate complex multivariate climate ensembles for any specific global region and time period. Second, as data arrive in blocks due to its online nature or storage limitations, the SG should be able to quickly adapt to incoming data and update itself. 

Statistically, the general idea behind existing SGs is to build a spatio-temporal stochastic model, possibly leveraging Gaussian processes (GPs) and vector autoregressive models (VARs), and then integrate various strategies to tackle the statistical problems posed by the data characteristics and practical demands. For instance, the TGH transformation is commonly included to Gaussianize climate data with high temporal resolution. To construct SGs for reanalysis data, two significant statistical problems should be addressed. First, efficiently modeling large multivariate geostatistical data. Strategies such as low-rank approximations \citep{FRK_intro}, covariance taperings \citep{taper2008}, composite likelihoods \citep{Katzfuss2021general}, and spatial partial differential equations \citep{bolin2023covariancebased}, which alleviate the computational challenges of large GPs, could be helpful. Second, efficiently updating the model as new data become available over time. Given the continuous accumulation of data, analyzing the entire dataset at once becomes impractical due to storage limitations and inefficient when timely emulations are required. The online updating strategy \citep{schifano2016online,benkeser2018online}, commonly applied to streaming data and dynamic environments, is particularly useful. It continuously updates the model as new data arrive using the developed updating formula, eliminating the need to store and process historical data or retrain the model from scratch.

This paper proposes an online SG (OSG) for the 3-hourly bivariate wind speed ERA5 ensembles over any given global region, which is able to efficiently generate emulations for the costly ERA5 ensembles with only model parameters to be stored. More importantly, we incorporate online updating into the SG, allowing data to enter the model sequentially in blocks and contribute to parameter updates. While maintaining performance comparable to the SG trained on the entire dataset, the OSG offers two additional advantages. First, it reduces computational resource demands during development, as only one block and essential quantities need to be stored. Second, it better adapts to the near real-time nature of ERA5 reanalysis data, enabling near real-time emulations. The key technique to enhance the efficiency and reduce storage requirements of the OSG is Slepian concentration. This method provides a set of data-independent basis functions that are optimally concentrated within the given region of interest. By leveraging these functions, we project the data into a lower-dimensional space, where their heavy-tail characteristics are modeled using the Tukey h transformation, and their dependence structure is captured by a VAR model. Another key element is the derivation of the online update formula for parameters, which typically combines cumulative estimates with those based on the current block, where the Lambert W function enables a closed-form update of parameters in the Tukey h transformation. Two applications of the proposed OSG are explored in our case studies. First, the OSG is used to address computational limitations by adjusting the block length according to available resources, leading to longer blocks and less frequent updates. Second, the OSG is designed to closely mimic the near real-time generation of ERA5 ensembles, rapidly adapting to incoming data with shorter blocks and more frequent updates. In both cases, the OSG performs comparably to the SG developed using the entire dataset at once.

The reminder of the paper is organized as follows. Section~\ref{sec:Data_Description} describes the ERA5 wind speed ensembles over the Arabian-Peninsula region. Section~\ref{sec:Methodology} presents the methodology for constructing the OSG, including an introduction to the key technique, Slepian concentration, and details on parameter estimation and updates within the OSG. Section~\ref{sec:Case_Study} demonstrates the development and performance of the OSG in two different scenarios. Finally, Section~\ref{sec:Discussion} summarizes this work and discusses potential extensions.

\section{ERA5 Reanalysis Data}
\label{sec:Data_Description}
The fifth generation ECMWF reanalysis data, known as ERA5 \citep{ERA52020}, is a product of the Copernicus Climate Change Service (C3S). It offers a detailed and comprehensive depiction of the global climate from 1940 onwards, integrating a big volume of observations with an advanced model and data assimilation system. ERA5 provides hourly data at a horizontal resolution of approximately $0.25^\circ$ ($17.29$ miles) and $137$ vertical levels, covering hundreds of atmospheric, ocean-wave and land-surface variables. It is daily updated and available up to five days prior to the present. Moreover, ERA5 includes $10$ ensemble members at lower spatial ($0.5^\circ$) and temporal ($3$-hourly) resolutions. These ensembles serve to quantify the relative uncertainties stemming from observations, sea surface temperature, and physical parametrizations of the model, identifying areas and periods of varying reliability. 

We choose ERA5 based on several key considerations. Firstly, ERA5 is highly representative. As a recently released global reanalysis dataset, it offers high spatial and temporal resolutions, a wide range of climate variables, and multiple ensembles for assessing data reliability. Widely utilized across diverse research domains, including studies into global and regional climate change, extreme events, and environmental monitoring, ERA5 stands as a trusted resource \citep{jiang2021evaluation,boettcher2023era5}. Secondly, ERA5 is readily accessible on the  \href{https://cds.climate.copernicus.eu/datasets/reanalysis-era5-single-levels?tab=download}{Copernicus Climate Data Store (CDS)} website. Through online portals and APIs, users can easily download data for specific regions and time periods. Lastly, ERA5 is computationally demanding. The generation, storage, and management of ERA5 and its ensembles demand advanced computational resources beyond the means of individual researchers.

\begin{figure}[!t]
    \centering
    \subfigure[Ensemble mean]{
    \label{Fig:subfig:Winduv_EnMean}
    \includegraphics[scale=0.49]{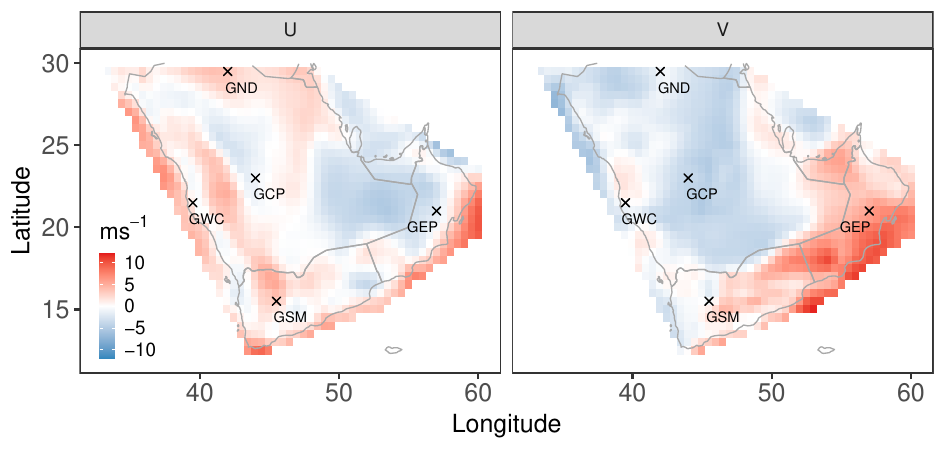}}
    \subfigure[Ensemble standard deviation]{
    \label{Fig:subfig:Winduv_EnSD}
    \includegraphics[scale=0.49]{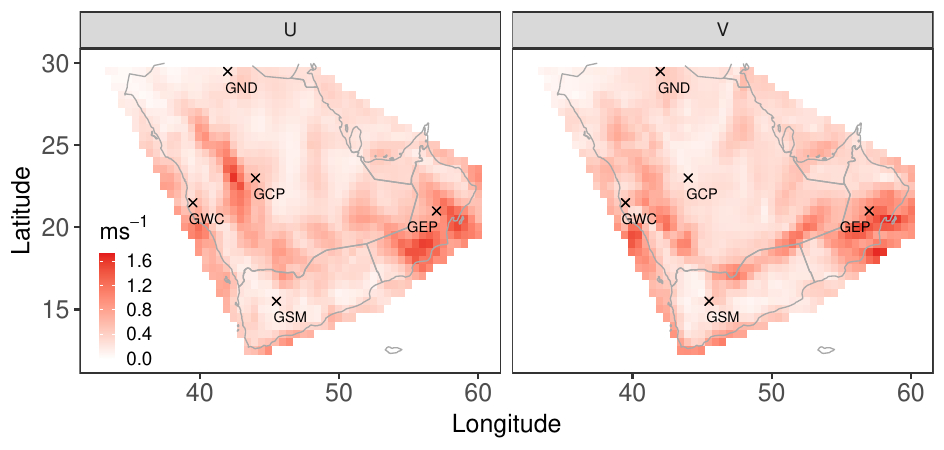}}\\
    \subfigure[Annual cycle]{
    \label{Fig:subfig:Winduv_AnnualCycle}
    \includegraphics[scale=0.49]{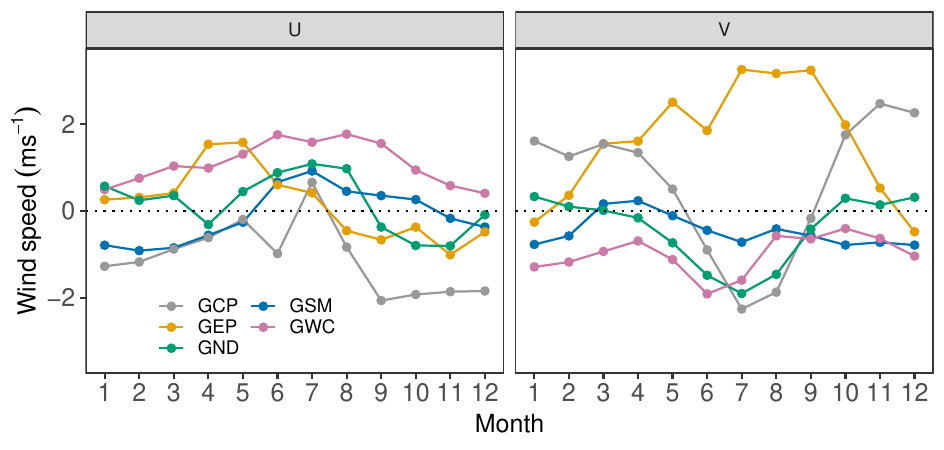}}
    \subfigure[Time series]{
    \label{Fig:subfig:Winduv_TimeSeries}
    \includegraphics[scale=0.49]{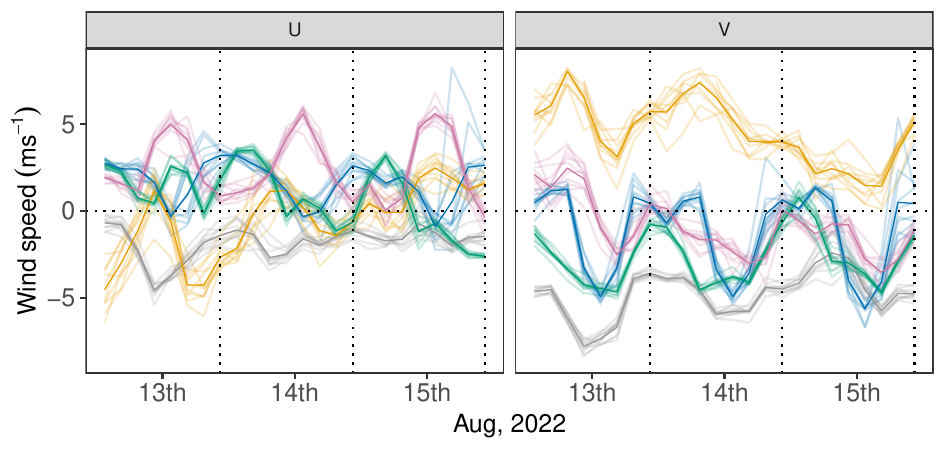}}\\
    \subfigure[Skewness]{
    \label{Fig:subfig:Winduv_Skew}
    \includegraphics[scale=0.49]{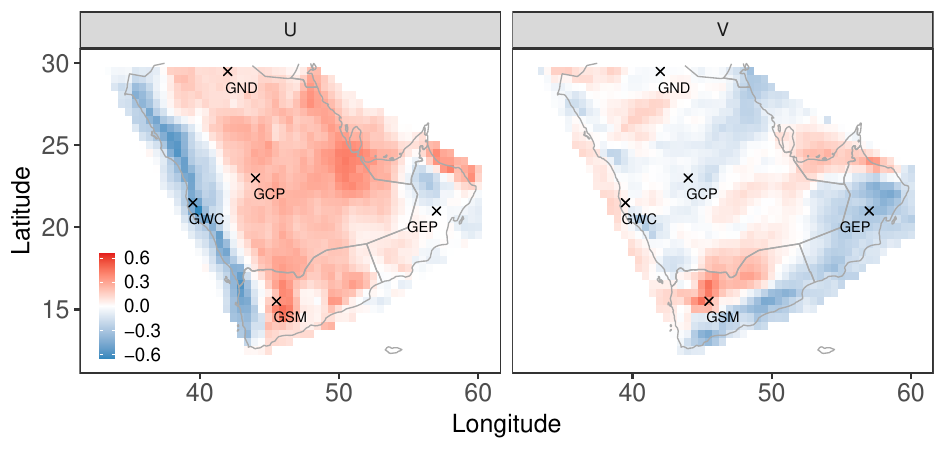}}
    \subfigure[Excess kurtosis]{
    \label{Fig:subfig:Winduv_Kurt}
    \includegraphics[scale=0.49]{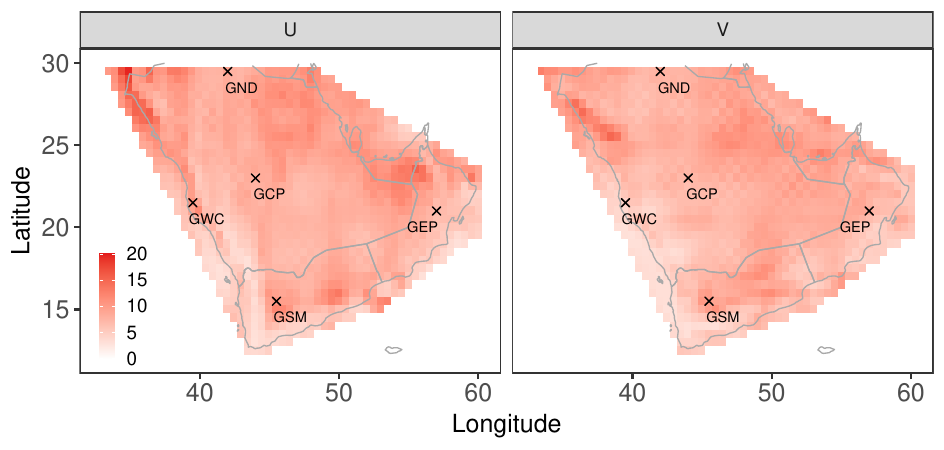}}\\
    \caption{Illustration of 10-meter U- and V-component wind speeds from ERA5 ensembles over the ARP. (a) and (b) show the ensemble means and standard deviations of wind speeds at 00:00 am on August 14, 2022, respectively. The gray lines indicate country borders, and five black crosses mark selected grid points: $\mathrm{GEP}=(21.0, 57.0)$, $\mathrm{GSM}=(15.5, 45.5)$, $\mathrm{GWC}=(21.5,39.5)$, $\mathrm{GND}=(29.5, 42.0)$, and $\mathrm{GCP}=(23.0, 44.0)$. (c) depicts annual cycles of wind speeds at the five grid points. (d) displays the 3-hourly wind speeds of all ensembles and their ensemble means at the five grid points from August 13th to 15th, 2022. (e) and (f) display the empirical skewness and excess kurtosis of wind speeds from all ensembles, respectively, after removing the ensemble mean.}
    \label{Fig:Demonstration}
\end{figure}

We consider the 10-meter U-component (eastward) and V-component (northward) wind speeds from ERA5 ensembles over the Arabian-Peninsula region (ARP) defined by \citet{essd-12-2959-2020}, spanning 2014--2023. As depicted in Fig.~S1, the ARP is surrounded by seas on three sides and exhibits diverse geographical features such as coastal plains, parallel mountain ranges along the western edge, a central plateau, and deserts. Let $y_{U,t}^{(r)}(L_i,l_j)$ and $y_{V,t}^{(r)}(L_i,l_j)$ represent U- and V-component wind speeds at ensemble $r$, $3$-hourly time point $t$ after the year 2013, and latitude$\times$longitude grid point $(L_i,l_j)$, respectively. Here, $r\in\Upsilon=\{1,\ldots,R\}$ with $R=10$, $t\in\mathcal{T}=\{1,\ldots,T\}$ with $T=8\times 365\times 10=29,200$, and $(L_i,l_j)\in\mathcal{G}_{\rm ARP}$. The number of grid points over the ARP is $|\mathcal{G}_{\rm ARP}|=1215$. The entire dataset encompasses approximately $710$ million data points. Fig.~\ref{Fig:Demonstration} illustrates various statistical characteristics of wind speeds across spatial and temporal domains, exhibiting the complex patterns and dynamics of 3-hourly wind speeds. 

Fig.~\ref{Fig:subfig:Winduv_EnMean} shows the ensemble means of wind speeds at a specific time point $t$, i.e., $\Bar{y}_{*,t}(L_i,l_j)=R^{-1}\sum_{r=1}^R y_{*,t}^{(r)}(L_i,l_j)$. Hereinafter, the symbol $*$ denotes that the  formula is applicable to both U- and V-components. Five selected grid points dispersed over the ARP are located at distinct geographical features: the eastern plain (GEP), southern mountains (GSM), western coast (GWC), northern desert (GND), and central plateau (GCP). Local geography significantly impacts wind speeds. For instance, nightly downslope winds from the western mountain ranges oppose the eastward winds from the Red Sea, moderating the eastward wind along the western coastline. The relatively flat terrain in the southeastern part of the ARP allows for stronger northward winds, which then counteract the southward winds from the north. Fig.~\ref{Fig:subfig:Winduv_EnSD} displays the ensemble standard deviations ${y}_{*,t}^{\mathrm{sd}}(L_i,l_j)=[R^{-1}\sum_{r=1}^R \{y_{*,t}^{(r)}(L_i,l_j)-\Bar{y}_{*,t}(L_i,l_j)\}^2]^{1/2}$, showing varying reliability across different regions. Both wind speeds show greater uncertainty near the two sides of the western mountain range, in the desert areas south of Saudi Arabia, and in the southeastern part of the ARP.

Fig.~\ref{Fig:subfig:Winduv_AnnualCycle} shows the annual cycles of wind speeds at five selected grid points, aggregated across ensembles and years. These curves primarily reflect the joint influence of climatic systems, seasonal variations, and geographical features on wind speed dynamics over the ARP, detailed in Section~S2. Additionally, these factors also influence the dependence both within individual wind speed components and between two components. For example, during summer, the Indian monsoon brings southeasterly winds from the Indian Ocean, significantly affecting the wind speeds on the southeastern ARP. Consequently, GEP exhibits notably strong northward winds, while westward winds are weaker, potentially obstructed by eastern mountain ranges. In other parts of the ARP, prevailing winds may be westerly or northwesterly. Fig.~\ref{Fig:subfig:Winduv_TimeSeries} zooms in further, depicting 3-hourly wind speeds over three days. The curves for GWC, GND, and GSM exhibit diurnal cycles, reflecting the influence of sea and land breezes, desert winds, and mountain-valley winds. Additionally, the data exhibit varying uncertainties at different time points. Explaining wind speed dynamics over the ARP remains challenging and necessitates further expert analysis. This study primarily focuses on the statistical aspects.

As discussed in the literature \citep{jeong2018reducing,tagle2019non}, high-temporal-resolution wind speed data often displays non-Gaussian characteristics. Fig.~\ref{Fig:Demonstration}\hyperref[Fig:subfig:Winduv_Skew]{(e)} and \hyperref[Fig:subfig:Winduv_Kurt]{(f)} depict the empirical skewness and excess kurtosis of $\{y_{*,t}^{(r)}(L_i,l_j)-\Bar{y}_{*,t}(L_i,l_j)\}_{t\in\mathcal{T}, r\in\Upsilon}$ at each grid point $(L_i,l_j)$. Generally, the skewness is not prominent. However, both wind speeds exhibit high kurtosis, indicating significant heavy-tailed distributions and more frequent outliers. 

\section{Methodology}
\label{sec:Methodology}
This section presents the development of an OSG in three steps. Section~\ref{sec:subsec:Slepian_concentration} introduces Slepian concentration, a key technique for capturing complex spatial structures and enhancing the efficiency of the proposed OSG. Section~\ref{sec:subsec:Spatio-temporal_modeling} constructs a standard SG for bivariate wind speed ERA5 ensembles over a specified global region and time interval. Based on this, Section~\ref{sec:subsec:Online_updating} provides the online updating procedures for parameters within the OSG.

Consider the data at time point $t$ on a closed region $\mathcal{R}$, denoted as $\{y_{*,t}^{(r)}(L_i,l_j)\}_{r\in\Upsilon,(L_i,l_j)\in\mathcal{G}_{\mathcal{R}}}$. While there is extensive literature on modeling large data using GPs, we need to address two additional considerations. First, since $\mathcal{R}$ lies on the sphere $\mathbb{S}^2$ and may cover several latitudes and longitudes, methods based on Euclidean distances are not suitable. Second, the data exhibits complex and non-stationary spatial patterns, making approaches that primarily rely on assumptions of stationarity or isotropy inadequate.

Therefore, we suggest employing low-rank approximation methods. Specifically, we assume that the data follow a spatial mixed effects model \citep{FRK2008}:
\vskip -20pt
\begin{equation}
    y_{*,t}^{(r)}(L_i,l_j)=m_{*,t}(L_i,l_j)+\sum_{\alpha=1}^A s_{*,t}^{(r)}(\alpha) g_{\alpha}(L_i,l_j)+\epsilon_{*,t}^{(r)}(L_i,l_j).
    \label{eq:LRK}
\end{equation}
The first term $m_{*,t}(L_i,l_j)$ represents the fixed mean trend shared across all ensembles. For each $(L_i,l_j)$, a parametric $m_{*,t}$ as in \citet{song2024efficient} is insufficient to capture the intrinsic, high-resolution wind speed dynamics. Hence, we directly set $\hat{m}_{*,t}=\Bar{y}_{*,t}$ and store the ensemble means  at all time and grid points, $\{\Bar{y}_{*,t}(L_i,l_j)\}_{t\in\mathcal{T},(L_i,l_j)\in\mathcal{G}_{\mathcal{R}}}$, which are readily available on the ERA5 website. With $A<|\mathcal{G}_{\mathcal{R}}|$, the second term provides a low-rank approximation of the random effect $z_{*,t}^{(r)}(L_i,l_j)=y_{*,t}^{(r)}(L_i,l_j)-m_{*,t}(L_i,l_j)$, where $g_{\alpha}$ denotes the $\alpha$th basis function and $s_{*,t}^{(r)}(\alpha)$ is the corresponding coefficient. The third term $\epsilon_{*,t}^{(r)}(L_i,l_j)$ captures the residual information and is assumed to be independent, following $\mathcal{N}(0,\sigma_{*,t}^2(L_i,l_j))$, where $\sigma_{*,t}^2(L_i,l_j)$ can be evaluated by $\hat{\sigma}_{*,t}^2(L_i,l_j)=R^{-1}\sum_{r=1}^R\{z_{*,t}^{(r)}(L_i,l_j)-\sum_{\alpha=1}^A s_{*,t}^{(r)}(\alpha) g_{\alpha}(L_i,l_j)\}^2$. 

\subsection{Slepian concentration problem on the sphere}
\label{sec:subsec:Slepian_concentration}
The selection of basis functions $\{g_{\alpha}(L_i,l_j)\}_{\alpha\in\mathcal{A}}$, where $\mathcal{A}=\{1,\ldots,A\}$ is crucial. Spherical harmonics \citep{jones1963stochastic} form a complete set of orthonormal basis functions defined in the Hilbert space of squared-integrable functions on $\mathbb{S}^2$. As shown in \citet{song2024efficient}, spherical harmonics efficiently represent global climate simulations. However, they may not stably represent functions on $\mathcal{R}$, as their lack of orthogonality and spatial concentration on the region introduces inherent ill-conditioning. The principal component analysis (PCA) is a powerful tool for reducing the dimension of a dataset while preserving as much variance as possible. The principal components form a set of orthogonal basis vectors on $\mathcal{R}$. However, these components are data-dependent. Principal components derived from U-component wind speeds may not adequately represent V-component wind speeds, and those based on one data block may not generalize well to subsequent blocks.

How can we obtain basis functions that are data-independent, orthogonal, and optimally concentrated on $\mathcal{R}$? Slepian concentration problem  \citep{slepian1961prolate,landau1961prolate,landau1962prolate,simons2006spatiospectral,Zubair2017,Zubair2020} offers a solution. 

We begin by introducing the necessary notations and concepts. For brevity, we assume $\mathbb{S}^2$ to be a unit sphere. Let $(\theta,\psi)$ represent a geographical point on $\mathbb{S}^2$, where $\theta=\pi/2-\pi L/180\in[0,\pi]$ and $\psi=\pi l/180\in[0,2\pi)$ are the reparameterizations of latitude $L$ and longitude $l$. Let $\mathcal{L}^2(\mathbb{S}^2)$ denote the Hilbert space of all real-valued, squared-integrable functions on $\mathbb{S}^2$, equipped with the inner product $\langle z,x \rangle=\int_{\mathbb{S}^2} z(\theta,\psi) x(\theta,\psi)\sin\theta\mathrm{d}\theta\mathrm{d}\psi$ and the norm (energy) $\|z\|\overset{\triangle}{=}\langle z,z\rangle^{1/2}$, where $\sin\theta\mathrm{d}\theta\mathrm{d}\psi$ is the differential area element on $\mathbb{S}^2$. Similarly, denote $\langle z,x\rangle_{\mathcal{R}}=\int_{\mathcal{R}} z(\theta,\psi)x(\theta,\psi)\sin\theta\mathrm{d}\theta\mathrm{d}\psi$ and $\|z\|_{\mathcal{R}}\overset{\triangle}{=}\langle z,x\rangle_{\mathcal{R}}^{1/2}$ as the inner product and the seminorm over the region $\mathcal{R}\subset \mathbb{S}^2$. Using real-valued spherical harmonics $\{h_q^m\}_{q=0,1,2,\ldots;m=-q,\ldots,q}$, as given in \citet{simons2006spatiospectral}, any $z\in \mathcal{L}^2(\mathbb{S}^2)$ can be represented in the spectral domain as $z(\theta,\psi)=\sum_{q=0}^{\infty}\sum_{m=-q}^q z_q^m h_q^m(\theta,\psi)$, where $z_q^m= \langle z,h_q^m \rangle$. In applications involving specific data, it is common to assume that $z$ is band-limited at some degree $Q$, such that all $z_q^m$ with $q\geq Q$ are zero. Then,
\vskip -20pt
\begin{equation}
    z(\theta,\psi)=\sum_{q=0}^{Q-1}\sum_{m=-q}^q z_q^m h_q^m(\theta,\psi).
    \label{eq:SHT}
\end{equation} 
The value of $Q$ depends on the data resolution, reflecting the level of detail in the dataset. The set $\mathcal{H}_{Q}=\{z: \langle z,h_q^m \rangle=0\text{ for }q\geq Q\}$ forms a $Q^2$-dimensional subspace of $\mathcal{L}^2(\mathbb{S}^2)$. 

Now, we turn to the Slepian concentration problem, following the idea outlined by \citet{simons2006spatiospectral}. Our goal is to find functions $g\in\mathcal{H}_Q$ that optimally concentrate their energy within the region $\mathcal{R}$. This is achieved by maximizing the ratio
\vskip -20pt
\begin{equation}
    \lambda=\frac{\|g\|_{\mathcal{R}}^2}{\|g\|^2}=\frac{\sum_{q=0}^{Q-1}\sum_{m=-q}^q\sum_{q'=0}^{Q-1}\sum_{m'=-q'}^{q'} g_q^m g_{q'}^{m'}C_{qm,q'm'}}{\sum_{q=0}^{Q-1}\sum_{m=-q}^q g_q^m g_q^m},
    \label{eq:Slepian1}
\end{equation}
where $C_{qm,q'm'}\overset{\triangle}{=}\int_{\mathcal{R}} h_q^m(\theta,\psi)h_{q'}^{m'}(\theta,\psi)\sin\theta\mathrm{d}\theta\mathrm{d}\psi$, and  $\lambda\in(0,1)$ measures the concentration. The second equality is obtained using \eqref{eq:SHT} and the orthonormality of spherical harmonics. Note that \eqref{eq:Slepian1} can be further expressed in matrix form as
\vskip -20pt
\begin{equation}
    \lambda=\frac{\mathbf{g}^\top\mathbf{C}\mathbf{g}}{\mathbf{g}^\top\mathbf{g}}, 
    \label{eq:Slepian2}
\end{equation}
where $\mathbf{g}=(g_0^0,g_{1}^{-1},g_1^0,g_1^1,g_2^{-2},\ldots,g_{Q-1}^{Q-1})^\top$ is a $Q^2$-dimensional vector, and $\mathbf{C}$ is a $Q^2\times Q^2$ real, symmetric, and positive definite matrix with elements $C_{qm,q'm'}$ indexed similarly to $\mathbf{g}$. The Slepian concentration problem then becomes solving the eigendecomposition $\mathbf{C}\mathbf{g}=\lambda\mathbf{g}$ in the spectral domain. 
Let $\lambda_{\alpha}$ denote the $\alpha$th largest eigenvalue of $\mathbf{C}$, and $\mathbf{g}_{\alpha}$ denote its normalized eigenvector, where $\alpha=1,\ldots,Q^2$. Plugging $\mathbf{g}_{\alpha}$ into \eqref{eq:SHT}, the resulting eigenfunction $g_{\alpha}(\theta,\psi)\in\mathcal{H}_Q$ in the spatial domain exhibits decreasing spatial concentration as $\alpha$ increases. With $\langle g_{\alpha}, g_{\beta}\rangle=\delta_{\alpha\beta}$ and $\langle g_{\alpha}, g_{\beta}\rangle_{\mathcal{R}}=\lambda_{\alpha}\delta_{\alpha\beta}$, these eigenfunctions are orthonormal on $\mathbb{S}^2$ and orthogonal on $\mathcal{R}$. Consequently, they form a complete set of bases for $\mathcal{H}_{Q}$, referred to as \textit{Slepian basis functions}. Although similar in their derivation, Slepian bases differ from principal components in that they are derived from $\mathbf{C}$, which depends only on the region and resolution, rather than from a data-driven covariance matrix.
\begin{figure}[!t]
\vspace{-20pt}
    \centering
    \subfigure[Slepian bases]{
    \label{fig:subfig:Slepian_Bases}
    \includegraphics[scale=1]{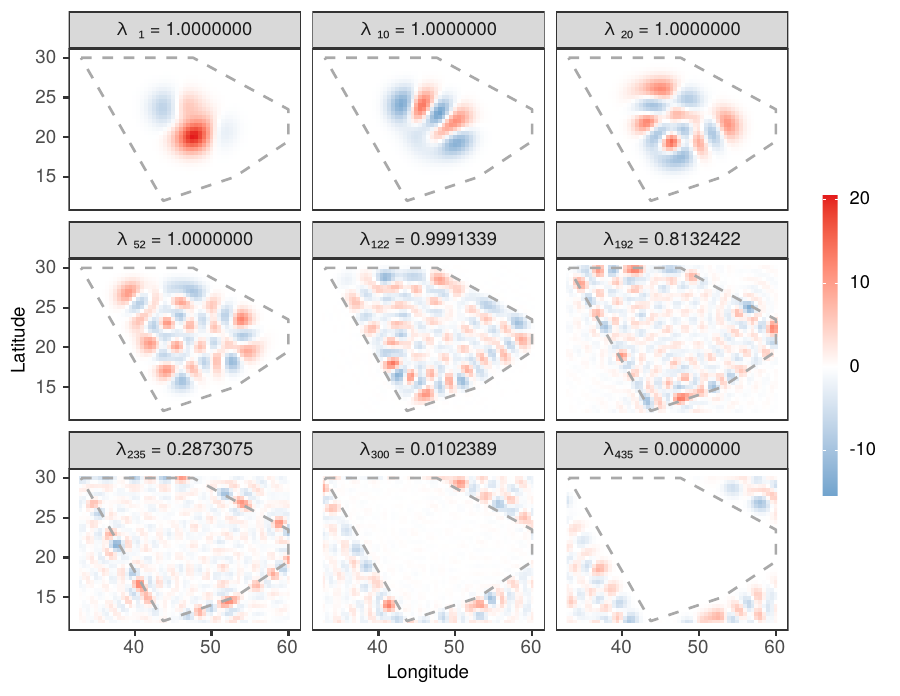}}\\
    \subfigure[Eigenvalues]{
    \hspace{-0.5in}
    \label{fig:subfig:Eigenvalues}
    \includegraphics[scale=0.65]{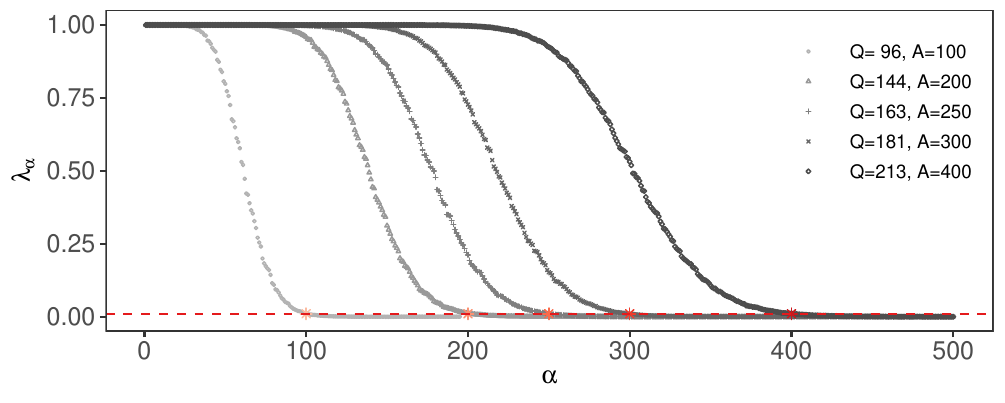}}
    \vspace{-10pt}
    \caption{Illustration of Slepian bases and eigenvalues. (a) shows Slepian bases for the ARP with $Q=181$ and their associated eigenvalues, which measure the degrees of concentration. The gray dotted line in each panel marks the boundary of the ARP. (b) shows the variation of eigenvalues $\lambda_{\alpha}$ for $\alpha\leq 500$ across different values of $Q$. The red dashed line marks the specified threshold of $0.01$.}
    \label{Fig:Slepian}
    \vspace{-10pt}
\end{figure}

Fig.~\ref{fig:subfig:Slepian_Bases} demonstrates several Slepian bases for the ARP with $Q=181$, along with their eigenvalues. These bases exhibit multiple resolutions. Those with $\lambda_{\alpha}=1$ are fully concentrated within the ARP. As $\lambda_{\alpha}$ decreases, the support of $g_{\alpha}$ gradually overflows the ARP and begins to fill its complementary set on the globe. When $\lambda_{\alpha}$ approaches zero, the corresponding $g_{\alpha}$ has negligible concentration on the ARP. Therefore, utilizing all $Q^2$ Slepian bases is often redundant. In the scenario where $\lambda_{\alpha}$ transitions sharply from nearly $1$ to nearly $0$, the sum of the eigenvalues $N=\sum_{\alpha=1}^{Q^2}\lambda_{\alpha}=\mathrm{tr}(\mathbf{C})$ approximates the number of Slepian bases that are well-concentrated on the region $\mathcal{R}$. This $N$, known as the Shannon number, depends only on the area of the region $\mathrm{area}(\mathcal{R})$ and $Q$, given by $N=\frac{\mathrm{area}(\mathcal{R})}{4\pi}Q^2$ \citep{simons2006spatiospectral}. For the example where $\mathcal{R}=\mathrm{ARP}$ and $Q=181$, $N$ is approximately $192$, with $g_{N}$ shown in the right panel of the middle row in Fig.~\ref{fig:subfig:Slepian_Bases}. The corresponding $\lambda_{N}$ is about $0.81$, indicating that Slepian bases with $\alpha>N$ still exhibit significant concentration on the ARP. In practice, one might select an $A$ between $N$ and $Q^2$ such that $\lambda_A\approx 0.01$, where $0.01$ is a suggested threshold that can be adjusted according to specific needs. As shown in the middle panel of the bottom row in Fig.~\ref{fig:subfig:Slepian_Bases}, $g_{300}$ has insignificant concentration on the ARP. Fig.~\ref{fig:subfig:Eigenvalues} demonstrates the first $500$ eigenvalues $\lambda_{\alpha}$ obtained using various $Q$ values, along with the corresponding $A$ values selected with a threshold of $0.01$. For a given region on the globe, $A$ is approximately proportional to $Q^2$, similar to the Shannon number $N$. Consequently, as data resolution increases, a larger $Q$ and more Slepian bases are necessary to capture finer details. 

\begin{figure}[!t]
\vspace{-20pt}
    \centering
    \subfigure[$A=100$ Slepian bases for \underline{both} variables]{
    \label{Fig:subfig:Slepian_Perform_A100}
    \includegraphics[scale=0.49]{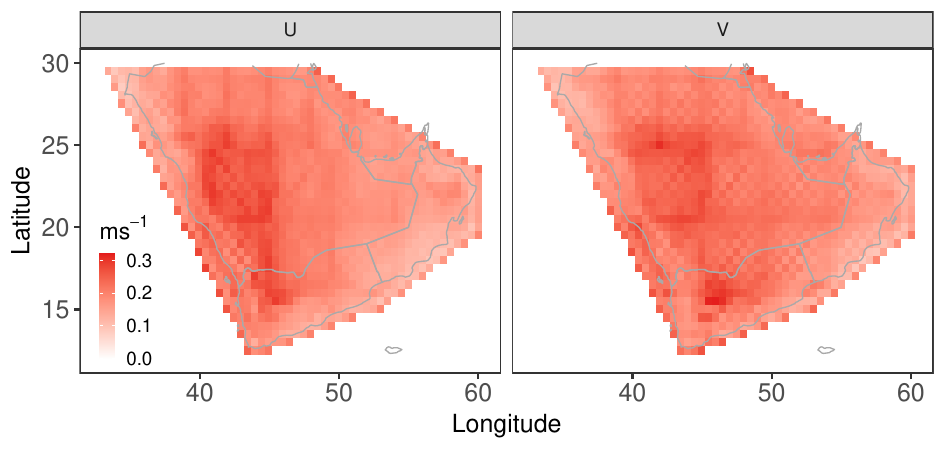}}
    \subfigure[$A=300$ Slepian bases for \underline{both} variables]{
    \label{Fig:subfig:Slepian_Perform_A300}
    \includegraphics[scale=0.49]{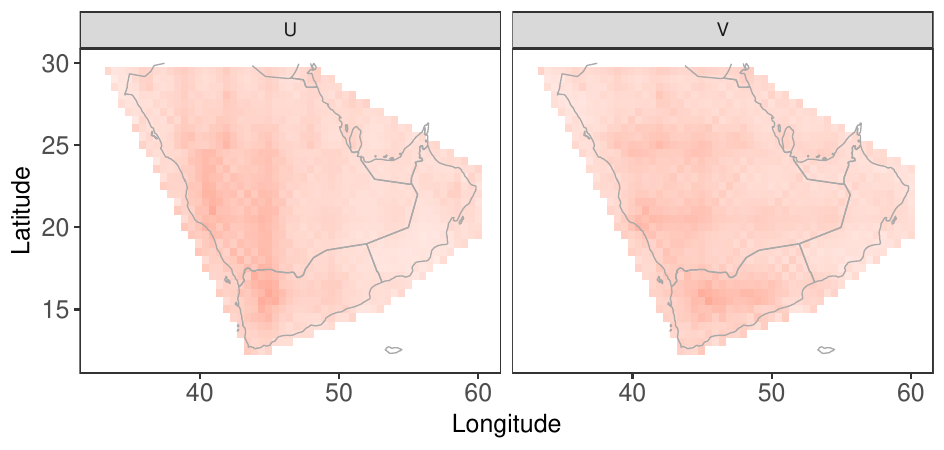}}\\
    \subfigure[$A=400$ Slepian bases for \underline{both} variables]{
    \label{Fig:subfig:Slepian_Perform_A400}
    \includegraphics[scale=0.49]{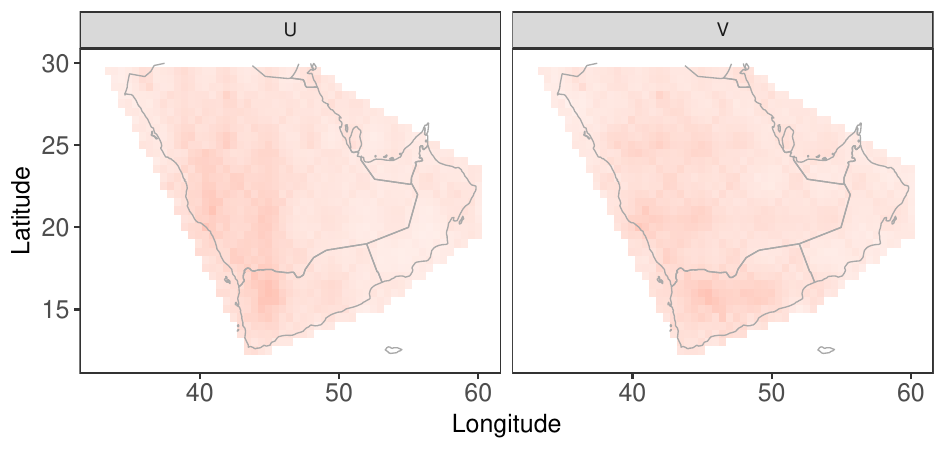}}
    \subfigure[$A=300$ principal components for \underline{each} variable]{
    \label{Fig:subfig:EOFs_Perform_A300}
    \includegraphics[scale=0.49]{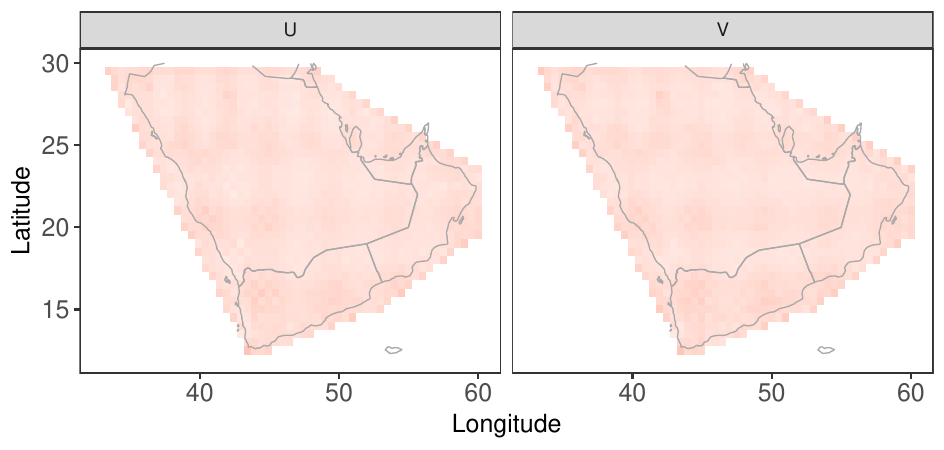}}\\
    \caption{Performance of Slepian bases. (a), (b) and (c) illustrate the approximation performance of $A=100$, $300$, and $400$ Slepian bases, respectively, measured by the rooted mean squared error $[(R|\mathcal{T}_t|)^{-1}\sum_{r=1}^R\sum_{t\in\mathcal{T}_t}\{\epsilon_{*,t}^{(r)}(L_i,l_j)\}^2]^{1/2}$, where $\mathcal{T}_t$ is a set of $1000$ randomly selected time points. (d) displays the approximation performance of $A=300$ principal components for each variable, necessitating a total of $600$ bases.}
    \label{Fig:Slepian_Performance}
\end{figure}

Fig.~\ref{Fig:Slepian_Performance}\hyperref[Fig:subfig:Slepian_Perform_A100]{(a)}--\hyperref[Fig:subfig:Slepian_Perform_A400]{(c)} illustrate the approximation performance of Slepian bases with increasing values of $A$ in representing wind speed ensembles across $1000$ random time points. Slepian bases effectively capture complex spatial structures, with performance improving significantly as $A$ grows. We also compare this to the performance of principal components, which provide optimal results and serve as a benchmark. Fig.~\ref{Fig:subfig:EOFs_Perform_A300} shows the performance of $300$ principal components for the U-component and another $300$ for the V-component wind speeds. By comparison, the $300$ Slepian bases shared by both variables perform comparably to principal components in most regions, although slightly higher errors are observed over the central plateau. We emphasize that the optimality of principal components is data-dependent, making them less suitable for our case. Unlike Slepian bases, principal components require additional storage since they cannot be derived independently of the data. Moreover, different variables and time periods may necessitate their respective principal components, leading to repeated calculations and increased storage demands. When examining dependencies between multiple variables, multivariate PCA further compounds the computational and storage burden.

\subsection{An SG for wind speed ERA5 ensembles over a fixed period}
\label{sec:subsec:Spatio-temporal_modeling}
This subsection details the procedures of constructing the SG and generating emulations for data over a fixed time period $\{y_{*,t}^{(r)}(L_i,l_j)\}_{t\in\mathcal{T}_0, r\in\Upsilon,(L_i,l_j)\in\mathcal{G}_{\mathcal{R}}}$, where $\mathcal{T}_0=\{1,\ldots,\tau_0\}$ with $\tau_0\leq T$. When $\tau_0 = T$, the SG is built using the entire dataset at once. First, we obtain the Slepian coefficients $\{s_{*,t}^{(r)}(\alpha)\}_{t\in\mathcal{T}_0, r\in\Upsilon, \alpha\in\mathcal{A}}$ by removing the ensemble means and applying Slepian concentration. In this step, $4\tau_0|\mathcal{G}_{\mathcal{R}}|$ parameters $\{\hat{m}_{*,t}(L_i,l_j)$, $\hat{\sigma}_{*,t}^2(L_i,l_j)\}_{t\in\mathcal{T}_0,  (L_i,l_j)\in\mathcal{G}_{\mathcal{R}}}$ in Model~\eqref{eq:LRK} are stored for use in the emulation procedure.

Second, we Gaussianize the coefficients, which inherit the heavy-tail characteristics from the random effect $z_{*,t}^{(r)}(L_i,l_j)$ shown in Fig.~\ref{Fig:subfig:Winduv_Kurt}. For each $\alpha$, we assume that the coefficients $\{s_{*,t}^{(r)}(\alpha)\}_{t\in\mathcal{T}_0, r\in\Upsilon}$ follow a scaled Tukey h distribution with parameters $\omega_{*,\alpha}>0$ and $h_{*,\alpha}\geq 0$: 
\vskip -20pt
\begin{equation}
    s_{*,t}^{(r)}(\alpha)=\omega_{*,\alpha}\Tilde{s}_{*,t}^{(r)}(\alpha)\exp[h_{*,\alpha}\{\Tilde{s}_{*,t}^{(r)}(\alpha)\}^2/2],
    \label{eq:TukeyH}
\end{equation}
where $\Tilde{s}_{*,t}^{(r)}(\alpha)$ follows a standard normal distribution. Using the Lambert W function $W(s)$, which satisfies $W(s)\exp\{W(s)\}=s$, \citet{goerg2015lambert} derived the inverse transformation of \eqref{eq:TukeyH} in closed form:
\vskip -30pt
\begin{equation}
    \Tilde{s}_{*,t}^{(r)}(\alpha)=W_{h_{*,\alpha}}(s_{*,t}^{(r)}(\alpha)/\omega_{*,\alpha}),
    \label{eq:InvTukeyH}
\end{equation}
where $W_h(s)=\mathrm{sgn}(s)\{W(hs^2)/h\}^{1/2}$ and $\mathrm{sgn}(s)$ denotes the sign of $s$. Additionally, he provided the analytic expressions for the second-order moment and kurtosis of $s_{*,t}^{(r)}(\alpha)$, given by  $\gamma_{*,\alpha}=\mathrm{E}\{s_{*,t}^{(r)}(\alpha)\}^2=\omega_{*,\alpha}^2(1-2h_{*,\alpha})^{-3/2}$ for $h_{*,\alpha}<1/2$, and $\kappa_{*,\alpha}=\mathrm{E}\{s_{*,t}^{(r)}(\alpha)\}^4/\gamma_{1,*(\alpha)}^2=3(1-2h_{*,\alpha})^3(1-4h_{*,\alpha})^{-5/2}$ for $h_{*,\alpha}<1/4$, respectively. These results lead to the estimates of $h_{*,\alpha}$ and $\omega_{*,\alpha}^2$ as follows:
\vskip -20pt
\begin{equation}
    \hat{h}_{*,\alpha}=\frac{1}{66}[\{66 \hat{\kappa}_{*,\alpha}-162\}^{1/2}-6]_+ \quad \text{and}\quad 
    \hat{\omega}_{*,\alpha}^2=\hat{\gamma}_{*,\alpha}(1-2\hat{h}_{*,\alpha})^{3/2}, 
    \label{eq:THestimate}
    \vspace{-10pt}
\end{equation}
where
\vskip -45pt
\begin{equation}
    \hat{\gamma}_{*,\alpha}=(R\tau_0)^{-1}\sum_{r=1}^R\sum_{t=1}^{\tau_0}\{s_{*,t}^{(r)}(\alpha)\}^2 \quad \text{and}\quad \hat{\kappa}_{*,\alpha}=(R\tau_0)^{-1}\sum_{r=1}^R\sum_{t=1}^{\tau_0}\{s_{*,t}^{(r)}(\alpha)\}^4/\hat{\gamma}_{*,\alpha}^2
    \label{eq:Gamma12Est}
    \vspace{-10pt}
\end{equation}
are empirical estimates of $\gamma_{*,\alpha}$ and $\kappa_{*,\alpha}$, respectively. Here, $[s]_+=\max(s,0)$, ensuring that $\hat{h}_{*,\alpha}>0$ when $\hat{\kappa}_{*,\alpha}>3$ and $\hat{h}_{*,\alpha}=0$ when $\hat{\kappa}_{*,\alpha}\leq 3$.
The first equality in \eqref{eq:THestimate} is derived from the Taylor expansion of $\kappa_{*,\alpha}$ around $h_{*,\alpha}=0$, specially $\kappa_{*,\alpha}=3+12h_{*,\alpha}+66h_{*,\alpha}^2+O(h_{*,\alpha}^3)$, with term $O(h_{*,\alpha}^3)$ being dropped. The derivation based on small $h_{*,\alpha}$ is appropriate for our real-world data, supported by the results in Figs.~\ref{Fig:subfig:h_alpha_Case1} and \ref{Fig:subfig:OmegaH_U_alpha_Case2}. By plugging the estimates from \eqref{eq:THestimate} into \eqref{eq:InvTukeyH}, we obtain the transformed coefficient $\{\Tilde{s}_{*,t}^{(r)}(\alpha)\}_{t\in\mathcal{T}_0, r\in\Upsilon, \alpha\in\mathcal{A}}$. In this step, $4A$ parameters $\{\hat{\gamma}_{*,\alpha},\hat{\kappa}_{*,\alpha}\}_{\alpha\in\mathcal{A}}$ or $\{\hat{\omega}_{*,\alpha},\hat{h}_{*,\alpha}\}_{\alpha\in\mathcal{A}}$ should be stored.

Finally, we model the dependence within the transformed coefficients. Let ${\tilde{\mathbf{s}}_t^{(r)}=[\Tilde{s}_{U,t}^{(r)}(1)}$, $\ldots,\Tilde{s}_{U,t}^{(r)}(A),\Tilde{s}_{V,t}^{(r)}(1),\ldots,\Tilde{s}_{V,t}^{(r)}(A)]^\top$ follow a vector auto-regressive model of order $P$ (VAR($P$)):
\vskip -20pt
\begin{equation}
\tilde{\mathbf{s}}_t^{(r)}=\sum_{p=1}^P\Phi_p \tilde{\mathbf{s}}_{t-p}^{(r)}+\bm\xi_t^{(r)},
\label{eq:VARP}
\end{equation}
where $t\in\mathcal{T}_{0/P}=\{P+1,\ldots,\tau_0\}$. Let $\tilde{\mathbf{S}}_t^{(r)\top}=[\tilde{\mathbf{s}}_{t-1}^{(r)\top},\ldots,\tilde{\mathbf{s}}_{t-P}^{(r)\top}]\in\mathbb{R}^{1\times 2AP}$ and $\Phi^\top=(\Phi_1,\ldots,\Phi_P)\in\mathbb{R}^{2A\times 2AP}$. Then, $\Phi$ can be estimated by 
\vskip -20pt
\begin{equation}
    \hat\Phi=\{\sum_{r=1}^R\sum_{t=P+1}^{\tau_0}\tilde{\mathbf{S}}_t^{(r)}\tilde{\mathbf{S}}_t^{(r)\top}\}^{-1}\sum_{r=1}^R\sum_{t=P+1}^{\tau_0}\tilde{\mathbf{S}}_t^{(r)}\tilde{\mathbf{s}}_t^{(r)\top},
    \label{eq:Phi_full}
\end{equation}
which captures the temporal evolution. To ensure the invertibility of the matrix in brackets, the length of the time period should satisfy $\tau_0\geq (2A+R)P/R$. The covariance matrix of residuals $\mathbf{K}=\mathrm{E}\{\bm\xi_t^{(r)}\bm\xi_t^{(r)\top}\}\in\mathbb{R}^{2A\times 2A}$ can then be evaluated by
\vskip -20pt
\begin{equation}
    \hat{\mathbf{K}}=\frac{1}{R(\tau_0-P)}\sum_{r=1}^R\sum_{t=P+1}^{\tau_0}\hat{\bm\xi}_t^{(r)}\hat{\bm\xi}_t^{(r)\top},
    \label{eq:K_full}
\end{equation}
where $\hat{\bm\xi}_t^{(r)}=\tilde{\mathbf{s}}_t^{(r)}-\sum_{p=1}^P\hat{\Phi}_p\tilde{\mathbf{s}}_{t-p}^{(r)}$. In this step, we store the matrices $\hat\Phi$ and $\hat{\mathbf{K}}$. 

The processes for constructing an SG and generating emulations for regional bivariate wind speed ERA5 ensembles over a fixed time period are summarized in Algorithms~S1 and S2, respectively. Compared to directly storing the dataset of size $2R\tau_0|\mathcal{G}_{\mathcal{R}}|$, the proposed SG requires only $4\tau_0|\mathcal{G}_{\mathcal{R}}|+4A+4(P+1)A^2$ parameters, with the three terms corresponding to the aforementioned steps. By setting $\tau_0=T$, we obtain the total number of parameters required for emulating the entire dataset.

\subsection{An OSG for wind speed ERA5 ensembles}
\label{sec:subsec:Online_updating}
Now, assume that the entire dataset consists of $B+1$ sequentially arriving blocks. Denote the initial data block as $\mathcal{Y}^{\{0\}}$, defined over $\mathcal{T}_0=\{1,\ldots,\tau_0\}$. For $b=1,\ldots,B$, the $(b+1)$th block $\mathcal{Y}^{\{b\}}=\{y_{*,t}^{(r)}(L_i,l_j)\}_{t\in\mathcal{T}_{b}, r\in\Upsilon, (L_i,l_j)\in\mathcal{G}_{\mathcal{R}}}$, with $\mathcal{T}_{b}=\{\tau_0+(b-1)\tau+1,\ldots,\tau_0+b\tau\}$, covers $\tau$ time points so that $\tau_0 + B\tau = T$. Denote $\mathcal{Y}^{[b]}=\{\mathcal{Y}^{\{0\}},\ldots, \mathcal{Y}^{\{b\}}\}$ as the cumulative data up to the $(b+1)$th block. Hereafter, the superscript $\{b\}$ will consistently indicate quantities derived specifically from the block $\mathcal{Y}^{\{b\}}$, whereas the superscript $[b]$ will refer to quantities based on the cumulative data $\mathcal{Y}^{[b]}$. This subsection details the development of an OSG.


Given the currently available block $\mathcal{Y}^{\{b\}}$, $\{m_{*,t}(L_i,l_j), \sigma_{*,t}^2(L_i,l_j)\}_{t\in \mathcal{T}_b; (L_i,l_j)\in\mathcal{G}_{\mathcal{R}}}$ (or subscripted by $\{b\}$) can be evaluated by calculating the ensemble means and performing Slepian concentration. These $2\tau|\mathcal{G}_{\mathcal{R}}|$ parameters depend only on the current block and can be stored directly without future updates. Our primary interest lies in deriving cumulative parameters $\{\hat{\gamma}_{*,\alpha}^{[b]},\hat{\kappa}_{*,\alpha}^{[b]}\}_{\alpha\in\mathcal{A}}$, $\hat{\Phi}^{[b]}$, and $\hat{\mathbf{U}}^{[b]}$ from the current block of Slepian coefficients $\mathcal{S}^{\{b\}}=\{s_{*,t}^{(r)}(\alpha)\}_{t\in\mathcal{T}_{b}, r\in\Upsilon, \alpha\in\mathcal{A}}$ and the cumulative parameters of $\mathcal{S}^{[b-1]}$ for  $b=1,\ldots,B$. 

First, we Gaussianize the current coefficient block $\mathcal{S}^{\{b\}}$ and update the parameters in the Tukey h transformation. For each $\alpha$, we calculate $\hat{\gamma}_{*,\alpha}^{\{b\}}$ and $\hat{\kappa}_{*,\alpha}^{\{b\}}$ with \eqref{eq:Gamma12Est}, $\hat{\omega}_{*,\alpha}^{\{b\}}$ and $\hat{h}_{*,\alpha}^{\{b\}}$ with \eqref{eq:THestimate}, and $\{\tilde{s}_{t}^{(r)}(\alpha)\}_{t\in\mathcal{T}_{b}, r\in\Upsilon}$ with \eqref{eq:InvTukeyH} in sequence. It can be shown that  $\hat{\gamma}_{*,\alpha}^{[b]}$ is a linear combination of the cumulative parameter $\hat{\gamma}_{*,\alpha}^{[b-1]}$ and the current parameter $\hat{\gamma}_{*,\alpha}^{\{b\}}$, with weights determined by the lengths and number of blocks. Specifically,
\vskip -20pt
\begin{equation}
    \hat{\gamma}_{*,\alpha}^{[b]}=\frac{\sum_{\iota=0}^{b-1}|\mathcal{T}_{\iota}|}{\sum_{\iota=0}^{b}|\mathcal{T}_{\iota}|}\hat{\gamma}_{*,\alpha}^{[b-1]}+\frac{|\mathcal{T}_{b}|}{\sum_{\iota=0}^{b}|\mathcal{T}_{\iota}|}\hat{\gamma}_{*,\alpha}^{\{b\}}=\frac{\tau_0+(b-1)\tau}{\tau_0+b\tau}\hat{\gamma}_{*,\alpha}^{[b-1]}+\frac{\tau}{\tau_0+b\tau}\hat{\gamma}_{*,\alpha}^{\{b\}},
    \label{eq:Gamma1_Update}
\end{equation}
where the first equality provides a generalized form that allows for varying block lengths. In the special case where $b=1$ and $\tau_0=\tau$, the cumulative parameter $\hat{\gamma}_{*,\alpha}^{[1]}$ for the first two blocks is simply the average of $\hat{\gamma}_{*,\alpha}^{[0]}=\hat{\gamma}_{*,\alpha}^{\{0\}}$ from the first block and $\hat{\gamma}_{*,\alpha}^{\{1\}}$ from the second block. Similarly, we can prove that
\vskip -15pt
\begin{equation}
\begin{aligned}
     \hat{\kappa}_{*,\alpha}^{[b]}&=\frac{\sum_{\iota=0}^{b-1}|\mathcal{T}_{\iota}|\left(\hat{\gamma}_{*,\alpha}^{[b-1]}\right)^2}{\sum_{\iota=0}^{b}|\mathcal{T}_{\iota}|\left(\hat{\gamma}_{*,\alpha}^{[b]}\right)^2}\hat{\kappa}_{*,\alpha}^{[b-1]}+\frac{|\mathcal{T}_b|\left(\hat{\gamma}_{*,\alpha}^{\{b\}}\right)^2}{\sum_{\iota=0}^{b}|\mathcal{T}_{\iota}|\left(\hat{\gamma}_{*,\alpha}^{[b]}\right)^2}\hat{\kappa}_{*,\alpha}^{\{b\}}\\
     &=
     \frac{\{\tau_0+(b-1)\tau\}\left(\hat{\gamma}_{*,\alpha}^{[b-1]}\right)^2}{(\tau_0+b\tau)\left(\hat{\gamma}_{*,\alpha}^{[b]}\right)^2}\hat{\kappa}_{*,\alpha}^{[b-1]}+\frac{\tau\left(\hat{\gamma}_{*,\alpha}^{\{b\}}\right)^2}{(\tau_0+b\tau)\left(\hat{\gamma}_{*,\alpha}^{[b]}\right)^2}\hat{\kappa}_{*,\alpha}^{\{b\}}.
\end{aligned}
\label{eq:Gamma2_Update}
\end{equation}
In the aforementioned special case, if we further assume that $\hat{\gamma}_{*,\alpha}^{\{0\}}=\hat{\gamma}_{*,\alpha}^{\{1\}}$, then $\hat{\kappa}_{*,\alpha}^{[1]}=\big(\hat{\kappa}_{*,\alpha}^{\{0\}}+\hat{\kappa}_{*,\alpha}^{\{1\}}\big)/2$. Denote $\hat{\gamma}_{*,\alpha}^{\rm full}$ and $\hat{\kappa}_{*,\alpha}^{\rm full}$ to be the estimates derived from the full dataset at once. We have $\hat{\gamma}_{*,\alpha}^{[B]}=\hat{\gamma}_{*,\alpha}^{\rm full}$ and $\hat{\kappa}_{*,\alpha}^{[B]}=\hat{\kappa}_{*,\alpha}^{\rm full}$, indicating that the online updating procedures do not result in any loss of accuracy in the parameter estimates for the Tukey h transformation. We also consider the scenario where the data exhibits both skewness and heavy tails, a common characteristic in climate data. In such cases, a TGH transformation is necessary, and the procedures for updating its parameters are detailed in Section~S3.2.

Second, we model the dependence within the current block of transformed coefficients $\tilde{\mathcal{S}}^{\{b\}}$ and update the matrices in the VAR($P$) model. By replacing $\mathcal{T}_0$ in \eqref{eq:Phi_full} and \eqref{eq:K_full} with $\mathcal{T}_b$, we can calculate the current parameters \small{$\hat{\Phi}^{\{b\}}=\left(\mathbf{X}^{\{b\}}\right)^{-}\sum_{r=1}^R\sum_{t\in\mathcal{T}_{b/P}}\mathbf{S}_t^{(r)}\mathbf{s}_t^{(r)\top}$} \normalsize and \small{$\hat{\mathbf{K}}^{\{b\}}=(R|\mathcal{T}_{b/P}|)^{-1}\{\sum_{r=1}^R\sum_{t\in\mathcal{T}_{b/P}}\mathbf{s}_t^{(r)}\mathbf{s}_t^{(r)\top}-\Phi^{\{b\}\top}\mathbf{X}^{\{b\}}\Phi^{\{b\}}\}$}, \normalsize where \small{$\mathbf{X}^{\{b\}}=\sum_{r=1}^R\sum_{t\in\mathcal{T}_{b/P}}\mathbf{S}_t^{(r)}\mathbf{S}_t^{(r)\top}$} \normalsize and \small{$\left(\mathbf{X}^{\{b\}}\right)^{-}$} \normalsize means a generalized inverse of \small{$\mathbf{X}^{\{b\}}$}. \normalsize Let \small $\mathbf{X}^{[b]}=\sum_{\iota=0}^b\mathbf{X}^{\{\iota\}}=\mathbf{X}^{[b-1]}+\mathbf{X}^{\{b\}}$. \normalsize The cumulative estimate of $\Phi$ can be represented as
\vskip -20pt
\begin{equation}
\hat{\Phi}^{[b]}=(\mathbf{X}^{[b-1]}+\mathbf{X}^{\{b\}})^{-1}(\mathbf{X}^{[b-1]}\hat{\Phi}^{[b-1]}+\mathbf{X}^{\{b\}}\hat{\Phi}^{\{b\}}). 
\label{eq:Phi_update}
\vspace{-10pt}
\end{equation}
Furthermore, the cumulative estimate of $\mathbf{K}$ satisfies 
\vskip -15pt
\begin{equation}
\begin{aligned}
R\sum_{\iota=0}^{b}|\mathcal{T}_{\iota/P}|\hat{\mathbf{K}}^{[b]}=&R\sum_{\iota=0}^{b-1}|\mathcal{T}_{\iota/P}|\hat{\mathbf{K}}^{[b-1]}+\hat{\Phi}^{[b-1]\top}\mathbf{X}^{[b-1]}\hat{\Phi}^{[b-1]}\\
&+R|\mathcal{T}_{b/P}|\hat{\mathbf{K}}^{\{b\}}
    +\hat{\Phi}^{\{b\}\top}\mathbf{X}^{\{b\}}\hat{\Phi}^{\{b\}}-\hat{\Phi}^{[b]\top}\mathbf{X}^{[b]}\hat{\Phi}^{[b]},
\end{aligned}
\label{eq:K_update}
\end{equation}
where the first two terms on the right-hand side come from the cumulative quantities up to the last block, the next two terms come from the current quantities, and the final term involves quantities just updated. 

The calculation of $\hat{\Phi}^{\{b\}}$ (and $\hat{\mathbf{K}}^{\{b\}}$) involves computing the generalized inverse of the matrix $\mathbf{X}^{\{b\}}$. When $\tau$ is small, $\mathbf{X}^{\{b\}}$ may not be full rank, potentially resulting in non-unique estimates for $\hat{\Phi}^{\{b\}}$ and $\hat{\mathbf{K}}^{\{b\}}$. However, this does not affect the uniqueness of the final estimates $\hat{\Phi}^{[B]}$ and $\hat{\mathbf{K}}^{[B]}$, as long as $\mathbf{X}^{[B]}$ is full rank, which can be guaranteed if the total number of time points is sufficient \citep{schifano2016online}. Therefore, the proposed OSG is applicable in cases with shorter blocks and more frequent updates. Let $\hat{\Phi}^{\rm full}$ and $\hat{\mathbf{K}}^{\rm full}$ represent the estimates based on the full dataset. In comparison, $\hat{\Phi}^{[B]}$ and $\hat{\mathbf{K}}^{[B]}$ experience a slight loss in accuracy due to the omission of the first $P$ time points from each block during updates. Consequently, as the number of blocks increases, this data loss accumulates. Nevertheless, these estimates remain good because the amount of data involved is sufficient to maintain accuracy.


In the frequentist framework, the online updating formulas illustrate how each data block influences various parameter estimates, both in magnitude and direction. In the Bayesian framework, these formulas utilize cumulative estimates as prior information, demonstrating how these priors combine with the current data block to yield posterior information, or the updated cumulative estimates. 

The entire workflow of developing an OSG for regional bivariate wind speed ERA5 ensembles is outlined in Algorithm~S3. Note that updating the matrices $\Phi$ and $\mathbf{K}$ requires an additional update and storage of the matrix $\mathbf{X}$, although it is not used for generating emulations. However, compared to storing and analyzing the full dataset of size $2RT|\mathcal{G}_{\mathcal{R}}|$, working with a single block of size $2R\max(\tau_0,\tau)|\mathcal{G}_{\mathcal{R}}|$ along with the matrix $\mathbf{X}$ of size $4A^2P^2$ significantly reduces the computational demands. Users can select blocks of appropriate length based on their objectives and available computing resources.

\section{Case Study}
\label{sec:Case_Study}
This section develops OSGs for the wind speed ERA5 ensembles described in Section~\ref{sec:Data_Description}, which involve two variables and cover a ten-year period over the ARP. The proposed OSG is applied to two scenarios. In the first, the OSG is built to reduce storage demands, allowing users to set longer blocks based on their system capacity, such as $\tau=365\times 8$. In the second, the OSG is designed to better mimic the data generation mechanism and enable near real-time emulations. The fast-update OSG requires short blocks, such as $\tau=7\times 8$. For both cases, we demonstrate the parameter estimates and updates within the OSGs and assess the emulation performance. For comparison, we also present the results of the SG built from the full dataset at once, referred to as FSG.

\subsection{Preliminaries}
\label{sec:subsec:Preliminary}
Before constructing the OSGs, some preliminary work is necessary. We introduce four indices to assess the performance of the proposed OSGs by comparing key statistical characteristics of the generated emulations, denoted as $\hat{y}_{*,t}^{(r)}(L_i,l_j)$, with those of the ERA5 ensembles. The first index 
\vskip -30pt
\begin{equation*}
    \mathrm{I}_{\rm uq,*}(L_i,l_j)=\frac{\mathrm{CRA}\{\hat y_{*,t}^{(1)}(L_i,l_j),\ldots,\hat y_{*,t}^{(R)}(L_i,l_j)\}}{\mathrm{CRA}\{y_{*,t}^{(1)}(L_i,l_j),\ldots,y_{*,t}^{(R)}(L_i,l_j)\}}
\end{equation*}
introduced by \citet{Huang'sEmulator}, quantifies the uncertainty of the emulations relative to the training  ensembles. Here, $\mathrm{CRA}\{y_{*,t}^{(1)}(L_i,l_j),\ldots,y_{*,t}^{(R)}(L_i,l_j)\}$ represents the central region area (CRA) of $R$ time series $\{y_{*,t}^{(r)}(L_i,l_j)\}_{r\in\Upsilon, t\in\mathcal{T}}$ at grid point $(L_i,l_j)$, which measures their interquantile range \citep{sun2011functional}. The closer the value is to $1$, the more the variability of the emulations aligns with that of the ERA5 ensembles. The second index, $\mathrm{I}_{\rm wdt,*}(L_i,l_j)$, follows \cite{song2024efficient} and measures the Wasserstein distance \citep{santambrogio2015optimal} between the empirical distributions of $\{y_{*,t}^{(r)}(L_i,l_j)\}_{r\in\Upsilon, t\in\mathcal{T}}$ and $\{\hat y_{*,t}^{(r)}(L_i,l_j)\}_{r\in\Upsilon, t\in\mathcal{T}}$ at grid point $(L_i,l_j)$. Similarly, the third index, $\mathrm{I}_{\rm wds,*}(t)$, quantifies the similarity between the empirical distributions of $\{y_{*,t}^{(r)}(L_i,l_j)\}_{r\in\Upsilon, (L_i,l_j)\in\mathcal{G}_{\mathcal{R}}}$ and $\{\hat y_{*,t}^{(r)}(L_i,l_j)\}_{r\in\Upsilon, (L_i,l_j)\in\mathcal{G}_{\mathcal{R}}}$ at time point $t$. 
For both indices, values closer to zero indicate better performance. Other measures can be applied depending on the specific purpose of the researchers. For example, to evaluate whether the emulations accurately reflect the unreliable regions identified in the ERA5 ensembles, we introduce the fourth index $\mathrm{I}_{\rm rq,*}(L_i,l_j)=\hat{y}_{*}^{\rm sd}(L_i,l_j)-y_{*}^{\rm sd}(L_i,l_j)=T^{-1}\sum_{t=1}^T\hat{y}_{*,t}^{\rm sd}(L_i,l_j)-T^{-1}\sum_{t=1}^Ty_{*,t}^{\rm sd}(L_i,l_j)$. A negative value of $\mathrm{I}_{\rm rq,*}(L_i,l_j)$ indicates that the emulations overestimate the reliability at grid point $(L_i,l_j)$, whereas a positive value suggests underestimation.

Using these indices and an initial data block of length $\tau_0=365\times 8$, we select the tuning parameter $A$ for the Slepian concentration and $P$ for the VAR model. First, we fix $P=2$ and vary $A$ from $100$ to $400$, constructing the corresponding SGs for the initial block. Fig.~S2(a)--(d) illustrates their performance, where both variables exhibit similar scales and patterns across all indices. The choice of $A$ has a small impact on the indices $\mathrm{I}_{\rm wdt,*}$ and $\mathrm{I}_{\rm wds,*}$, whereas significantly affects $\mathrm{I}_{\rm uq,*}$ and $\mathrm{I}_{\rm rq,*}$. Therefore, we present boxplots of the combined indices $\mathrm{I}_{\rm uq}(L_i,l_j)=\{\mathrm{I}_{\rm uq,U}(L_i,l_j)+\mathrm{I}_{\rm uq,V}(L_i,l_j)\}/2$ and $\mathrm{I}_{\rm rq}(L_i,l_j)=\{\mathrm{I}_{\rm rq,U}(L_i,l_j)+\mathrm{I}_{\rm rq,V}(L_i,l_j)\}/2$ in Fig.~S2(e) and (f). By minimizing $\sum_{(L_i,l_j)\in\mathcal{G}_{\mathrm{ARP}}}|\mathrm{I}_{\rm uq}(L_i,l_j)-1|$ and  $\sum_{(L_i,l_j)\in\mathcal{G}_{\mathrm{ARP}}}|\mathrm{I}_{\rm rq}(L_i,l_j)|$, we choose $A=300$.

Next, we determine the parameter $P$ by referencing to the partial autocorrelation \citep[PAC,][]{box2013box}. For $p=1,\ldots,4$, we evaluate the $p$th order PAC between the transformed coefficient time series $\tilde{s}_{*,t}^{(r)}(\alpha_1)$ and $\tilde{s}_{*,t-p}^{(r)}(\alpha_2)$ for any $\alpha_1,\alpha_2\in\mathcal{A}$, corresponding to the $(\alpha_1,\alpha_2)$ element of the matrix $\mathbf{P}_{*,p}\in\mathbb{R}^{A\times A}$. Additionally, we compute the matrices $\mathbf{P}_{UV,p}$ and $\mathbf{P}_{VU,p}$, which consist of the $p$th order PAC between $\tilde{s}_{U,t}^{(r)}(\alpha_1)$ and $\tilde{s}_{V,t-p}^{(r)}(\alpha_2)$, and $\tilde{s}_{V,t}^{(r)}(\alpha_1)$ and   $\tilde{s}_{U,t}^{(r)}(\alpha_1)$, respectively. Fig.~S3 illustrates these matrices. For $p=1$ and $2$, all matrices exhibit similar patterns, with values in $\mathbf{P}_{UV,p}$ and $\mathbf{P}_{VU,p}$ being considerably smaller than those in $\mathbf{P}_{U,p}$ and $\mathbf{P}_{V,p}$. The diagonal elements and those nearby show significant correlations, indicating dependence among various coefficients both within and between variables. However, this pattern becomes less evident after $p=2$. Considering these results along with storage requirements, we choose $P=2$. Further details regarding the selection of tuning parameters can be found in Section~S4.1 of the Supplementary Materials.

\subsection{Scenario 1: OSG with long blocks}
\label{sec:subseq:Case1}
Now, we consider the first scenario, where the available computational resources are insufficient to store and analyze ten years of ERA5 ensembles simultaneously, requiring us to process the data sequentially in blocks. We set $\tau_0=\tau=365\times 8$ and $B=9$, and construct an OSG capable of emulating the entire dataset while working within these computational constraints. Each time a new data block arrives, the parameters $m_{*,t}(L_i,l_j)$ and $\sigma_{*,t}(L_i,l_j)$ are evaluated, and the cumulative estimates of the parameters in the Tukey h transformation and the VAR($2$) model are updated.

 \begin{figure}[!b]
    \centering
    \subfigure[$\hat{\omega}_{*,\alpha}^{[0]}$ and $\hat{\omega}_{*,\alpha}^{[B]}$]{
    \label{Fig:subfig:Omega_alpha_Case1}
    \includegraphics[scale=0.7]{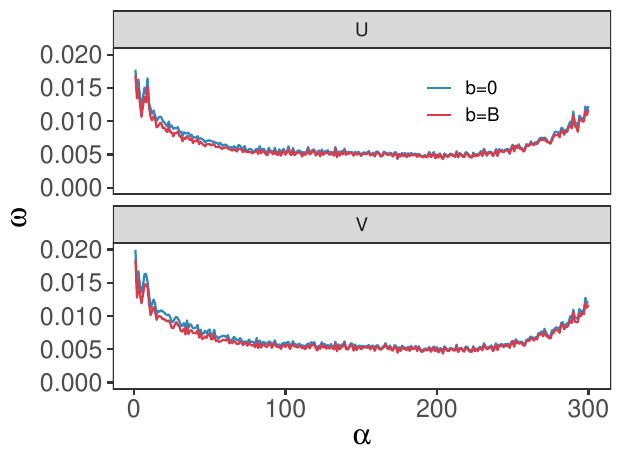}}
    \subfigure[$\hat{h}_{*,\alpha}^{[0]}$ and $\hat{h}_{*,\alpha}^{[B]}$]{
    \label{Fig:subfig:h_alpha_Case1}
    \includegraphics[scale=0.7]{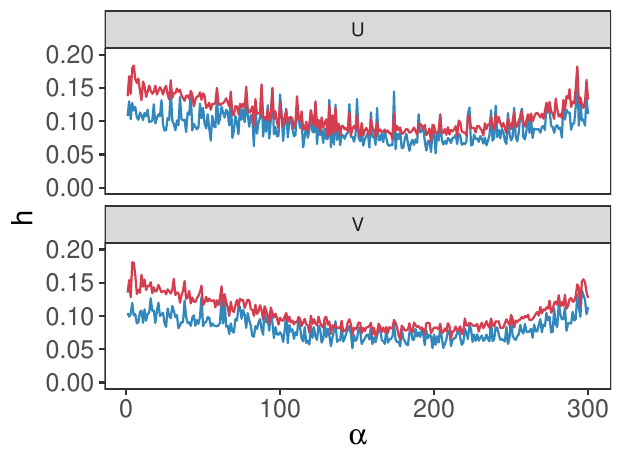}}\\
    \subfigure[$\hat{\omega}_{*,1}^{[b]}$ and $\hat{\omega}_{*,151}^{[b]}$]{
    \label{Fig:subfig:omega_b_Case1}
    \includegraphics[scale=0.7]{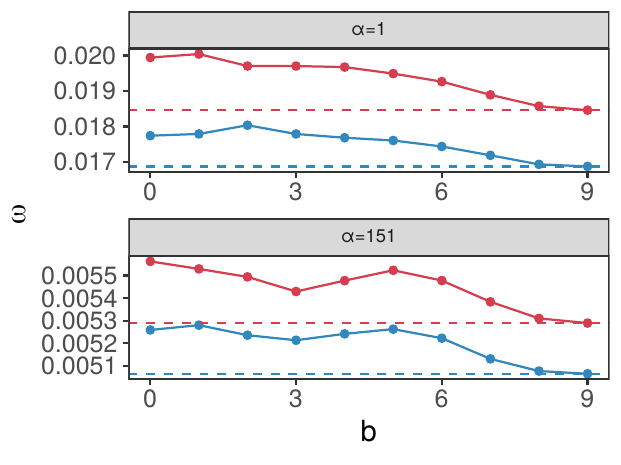}}
    \subfigure[$\hat{h}_{*,1}^{[b]}$ and $\hat{h}_{*,151}^{[b]}$]{
    \label{Fig:subfig:h_b_Case1}
    \includegraphics[scale=0.7]{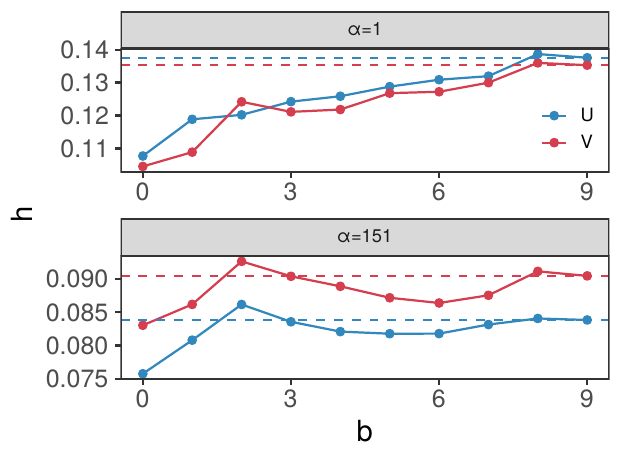}}
    \vspace{-10pt}
    \caption{Estimates and updates of parameters $\omega_{*,\alpha}$ and $h_{*,\alpha}$ in the Tukey h transformation. (a) and (b) illustrate the cumulative estimates of $\{\omega_{*,\alpha}\}_{\alpha\in\mathcal{A}}$ and $\{h_{*,\alpha}\}_{\alpha\in\mathcal{A}}$, respectively, up to the initial and the final blocks. (c) and (d) illustrate the updates of $\omega_{*,\alpha}^{[b]}$ and $h_{*,\alpha}^{[b]}$, respectively, over successive blocks, for $\alpha=1$ and $151$. The dashed lines represent the estimates derived from the complete ten years of data.}
    \label{Fig:Converge_Tukeyh_Case1}
    \vspace{-30pt}
\end{figure}

Fig.~\ref{Fig:Converge_Tukeyh_Case1} illustrates the estimates and updates of the parameters $\omega_{*,\alpha}$ and $h_{*,\alpha}$. Fix $b=0$ and $B$, we can see how $\hat\omega_{*,\alpha}^{[b]}$ and $\hat{h}_{*,\alpha}^{[b]}$ change with $\alpha$ from Fig.~\ref{Fig:Converge_Tukeyh_Case1}\hyperref[Fig:subfig:Omega_alpha_Case1]{(a)} and \hyperref[Fig:subfig:h_alpha_Case1]{(b)}, respectively. As $\alpha$ increases, both parameters initially decrease and then rise, indicating that the Slepian coefficients $s_{*,t}^{(r)}(\alpha)$ at both small and large $\alpha$ exhibit greater variation and more pronounced heavy tails. For smaller $\alpha$, the terms $s_{*,t}^{(r)}g_{\alpha}(L_i,l_j)$ tend to capture larger-scale, lower-frequency variations concentrated within the ARP, which reflect key characteristics of the data. For larger $\alpha$, these terms capture variations near the boundary of the ARP, where wind speed dynamics are more complex. 

Fig.~\ref{Fig:Converge_Tukeyh_Case1}\hyperref[Fig:subfig:omega_b_Case1]{(c)} and \hyperref[Fig:subfig:h_b_Case1]{(d)} demonstrate the updates of $\hat{\omega}_{*,\alpha}^{[b]}$ and $\hat{h}_{*,\alpha}^{[b]}$ as $b$ increases, along with the estimates $\hat{\omega}_{*,\alpha}^{\rm full}$ and $\hat{h}_{*,\alpha}^{\rm full}$ obtained from the full dataset for comparison. We set $\alpha$ to $1$ and $151$, corresponding to different scales of variance and degrees of heavy tails in the Slepian coefficients. All cumulative estimates converge stably to $\hat{\omega}_{*,\alpha}^{\rm full}$ and $\hat{h}_{*,\alpha}^{\rm full}$ as $b$ increases, with each update reflecting the influence of the current data block on the parameter estimates. In Fig.~\ref{Fig:subfig:h_b_Case1}, $\hat{h}_{*,1}^{[b]}$ continues to increase with $b$, suggesting that each newly arriving block  introduces additional extreme values to the Slepian coefficients at $\alpha=1$, which exhibit more pronounced heavy-tail characteristics. The update of $\hat{h}_{*,1}^{[b]}$ from $b=8$ to $b=9$ is relatively small, indicating that the information within the final block has already been captured by the preceding ones. For Slepian coefficients at $\alpha=151$, which are slightly heavy-tailed, the estimate $\hat{h}_{*,151}^{[b]}$ gradually stabilizes around $\hat{h}_{*,151}^{\rm full}$ as additional data is incorporated. Additionally, the patterns of $\hat{\omega}_{*,\alpha}^{[b]}$ are negatively correlated with those of $\hat{h}_{*,\alpha}^{[b]}$, consistent with the equality in~\eqref{eq:THestimate}. An increase in $\hat{h}_{*,\alpha}^{[b]}$ always corresponds to a decrease in $\hat{\omega}_{*,\alpha}^{[b]}$. 

\begin{figure}[!t]
    \vspace{-30pt}
    \centering
    \subfigure[$\hat{\Phi}_1^{[b]}$ and $\hat{\Phi}_1^{\rm full}$]{
    \hspace{-10pt}
    \label{Fig:subfig:Phi1_Case1}
    \includegraphics[scale=0.48]{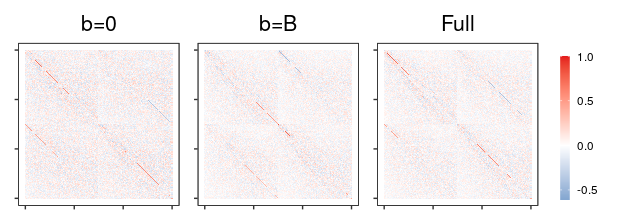}}
    \subfigure[$\hat{\mathbf{K}}^{[b]}$ and $\hat{\mathbf{K}}^{\rm full}$]{
    \label{Fig:subfig:K_Case1}
    \includegraphics[scale=0.48]{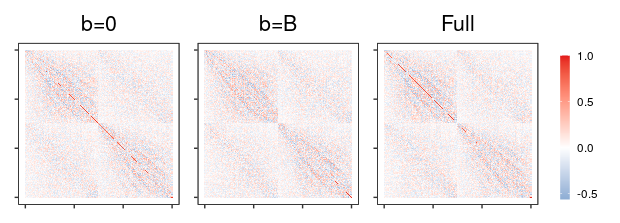}}
    \vspace{-10pt}
    \caption{Estimates and updates of the matrices $\Phi_1$ and $\mathbf{K}$ in the VAR($2$) model. (a) and (b) display the rescaled cumulative estimates of $\Phi$ and $\mathbf{K}$, respectively, after the first block and the final block, along with their rescaled estimates obtained from the full ten-year data. All elements have been rescaled using the function $f(x)=\mathrm{sign}(x)\sqrt{|x|}$ for clearer visualization.}
    \label{Fig:Converge_VAR2_Case1}
\end{figure}

Fig.~\ref{Fig:Converge_VAR2_Case1} illustrates the cumulative estimates of matrices $\Phi_1$ and $\mathbf{K}$ in the VAR($2$) model, along with $\hat{\Phi}_1^{\rm full}$ and $\hat{\mathbf{K}}^{\rm full}$. Each matrix comprises four sub-matrices representing the dependence between transformed coefficients within U (top-left), V (bottom-right), and across U and V (top-right and bottom-left). In each sub-matrix of $\hat\Phi_{1}^{[b]}$, the diagonal elements show significant values, meaning strong temporal dependence between coefficients at the same index $\alpha$, both within and across variables. Additionally, elements near the diagonal also exhibit notable values, reflecting the temporal dependence between coefficients at different indices. Compared to $\hat{\Phi}_1^{[0]}$, $\hat{\Phi}_1^{[B]}$ reveals clearer patterns and closely resembles $\hat{\Phi}_1^{\rm full}$. Estimates for $\Phi_2$ are provided in Fig.~S4 of the Supplementary Materials and lead to similar conclusions as those for $\Phi_1$. 

Taking the index $\alpha$ as the location of the coefficients $\tilde{s}_{*,t}^{(r)}(\alpha)$, Fig.~\ref{Fig:subfig:K_Case1} shows the ``spatial" dependence among the coefficients. Coefficients with closer indices exhibit stronger dependence, which results in a dark band along the diagonals of the sub-matrices in $\hat{\mathbf{K}}^{[b]}$ and $\hat{\mathbf{K}}^{\rm full}$. As $\alpha$ increases, this band widens, suggesting that the ``spatial" dependence is ``non-stationary". When $\alpha$ is large, more Slepian bases with similar resolution levels jointly cover the ARP region, which may lead to dependence among their coefficients. Comparing $\hat{\mathbf{K}}^{[0]}$ with $\hat{\mathbf{K}}^{[B]}$ and $\hat{\mathbf{K}}^{\rm full}$ reveals small updates, meaning that the initial block provides sufficient information to model the ``spatial" dependence. 

Denote $\mathrm{RFD}(\mathbf{A},\mathbf{B})=\|\mathbf{A}-\mathbf{B}\|_F/\|\mathbf{B}\|_F$ to be the relative Frobenius distance between matrices $\mathbf{A}$ and $\mathbf{B}$. Numerically, we calculate the RFDs between cumulative estimates and the estimates based on the full dataset, as shown in Table~S1 of the Supplementary Materials. The RFDs for all cumulative estimates gradually decreases as more data blocks are incorporated, while the decrease for $\hat{\mathbf{K}}^{[b]}$ is smaller and slower. This is consistent with our previous observations.

\begin{figure}[!t]
\vspace{-20pt}
    \centering
    \subfigure[$\mathrm{I}_{\rm uq,U}$]{
    \label{Fig:subfig:Iuqu_Online_Case1}
    \includegraphics[scale=0.43]{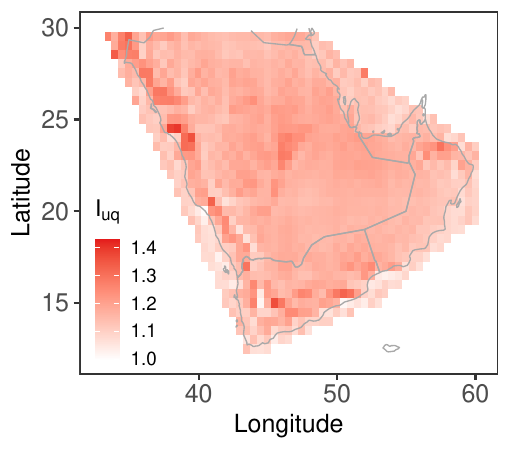}}
    \subfigure[$\mathrm{I}_{\rm wdt,U}$]{
    \label{Fig:subfig:Iwdtu_Online_Case1}
    \includegraphics[scale=0.43]{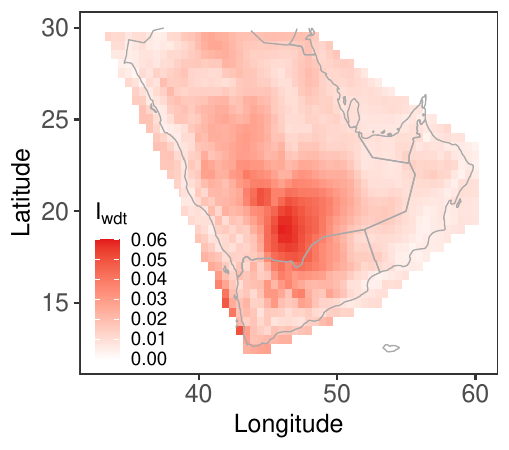}}
    \subfigure[$\mathrm{I}_{\rm wds,U}$]{
    \label{Fig:subfig:Iwdsu_Online_Case1}
    \includegraphics[scale=0.43]{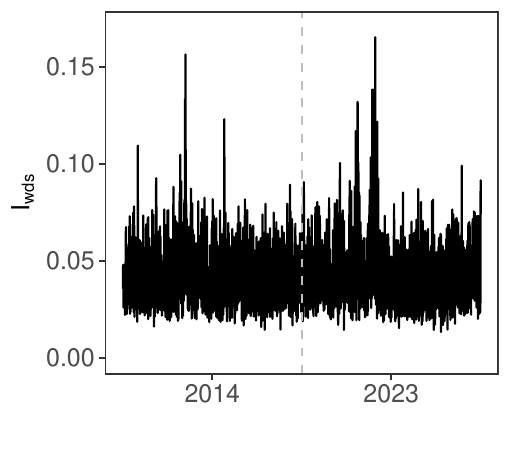}}
    \subfigure[$\mathrm{I}_{\rm rq,U}$]{
    \label{Fig:subfig:Irqu_Online_Case1}
    \includegraphics[scale=0.43]{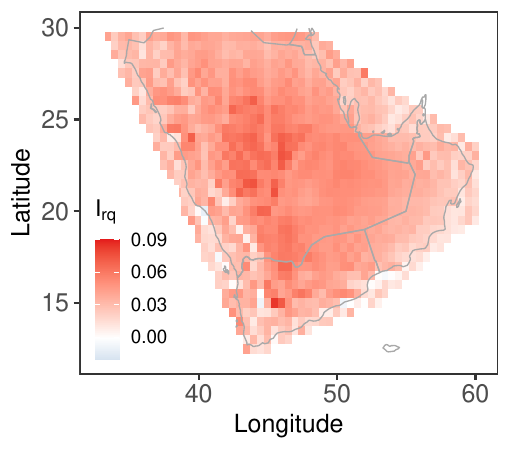}}
    \vspace{-10pt}
    \caption{Performance of the OSG constructed for the first scenario. Various metrics are presented to assess the U-component emulations. (c) provides partial results.}
    \label{Fig:Assessment_Case1}
\end{figure}

Given cumulative estimates up to the final block, we construct an OSG for the first scenario and generate emulations for all time points. Since these estimates closely match those obtained from the full ten-year dataset, the OSG closely resembles the FSG. In Fig.~\ref{Fig:Assessment_Case1}, we illustrate the performance of the OSG using the four aforementioned metrics and compare it to the performance of the FSG. To save space, only the results for the U component are shown, while those for V component, which lead to similar conclusions, are provided in Fig.~S5 of the Supplementary Materials. The magnitudes of the metrics indicate that the performance of the OSG is excellent across all measures and closely matches that of the FSG. These various metrics evaluate the emulations from different perspectives. From Figs.~\ref{Fig:subfig:Iuqu_Online_Case1}, \ref{Fig:subfig:Irqu_Online_Case1}, S5(a), and S5(d), both $\mathrm{I}_{\rm uq,*}$ and $\mathrm{I}_{\rm rq,*}$ are higher in the mountain ranges. Referring to Figs.~\ref{Fig:subfig:Winduv_EnSD} and \ref{Fig:subfig:Uncertaintyu_original}, the standard deviations in these regions are relatively lower, making it more challenging for the emulations to closely resemble the EAR5 ensembles. As shown in Figs.~\ref{Fig:subfig:Iwdtu_Online_Case1} and S5(b), emulations in regions where the heavy-tail characteristics of data are less pronounced exhibit suboptimal performance. It aligns with the observations in the preliminary work, reflecting an over-fitting of the heavy-tail characteristics with $A=300$ Slepian bases. Consequently, several high $\mathrm{I}_{\rm wds,*}$ values are also observed in Figs.~\ref{Fig:subfig:Iwdsu_Online_Case1} and S5(c). From Figs.~\ref{Fig:subfig:Irqu_Online_Case1} and S5(d), regions with greater uncertainty tend to exhibit higher $\mathrm{I}_{\rm rq}$ values. Furthermore,  Fig.~\ref{Fig:Assessment_Case2}\hyperref[Fig:subfig:Boxplot_Iuqu_Case2]{(e)}--\hyperref[Fig:subfig:Boxplot_Irqu_Case2]{(h)} demonstrate that the OSG performs comparably to the FSG.

\subsection{Scenario 2: OSG with short blocks}
\label{sec:subseq:Case2}
In the second scenario, we aim to develop a fast-update OSG to better mimic the generation mechanism of the ERA5 ensembles. To achieve this, we set $\tau_0=31\times 8$, $\tau=7\times 8$, and $B=517$, enabling weekly updates for the OSG. The initial block is longer than subsequent blocks to ensure unique estimates for $\mathbf{X}^{[b]}$ and $\hat{\mathbf{K}}^{[b]}$ for all $b=0,\ldots,B$. Compared to the OSG in the first scenario (OSG-Long), this OSG (OSG-Short) adapts more efficiently to new data and reduces the storage with each update. However, the parameter updates are less stable, and estimates of parameters $\Phi_1$, $\Phi_2$, and $\mathbf{K}$ are less accurate than in the previous scenario. 

\begin{figure}[!b]
\vspace{-10pt}
    \centering
    \subfigure[$\hat{\omega}_{U,\alpha}^{[0]}$, $\hat{\omega}_{U,\alpha}^{[B]}$, $\hat{h}_{U,\alpha}^{[0]}$, and $\hat{h}_{U,\alpha}^{[B]}$]{
    \label{Fig:subfig:OmegaH_U_alpha_Case2}
    \includegraphics[scale=0.7]{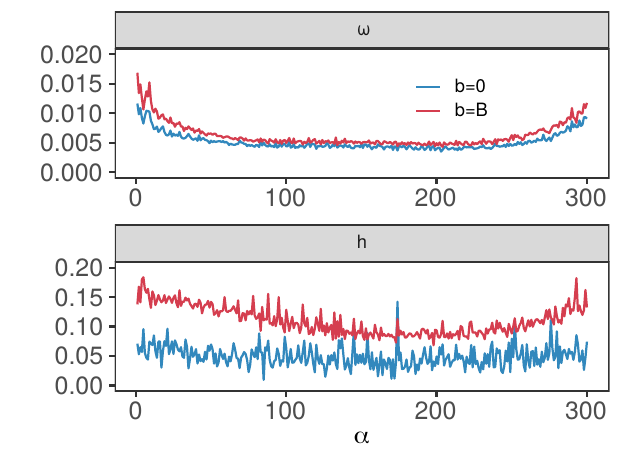}}
    \subfigure[$\hat{\omega}_{*,1}^{[b]}$ and $\hat{h}_{*,1}^{[b]}$]{
    \label{Fig:subfig:OmegaH_1_b_Case2}
    \includegraphics[scale=0.7]{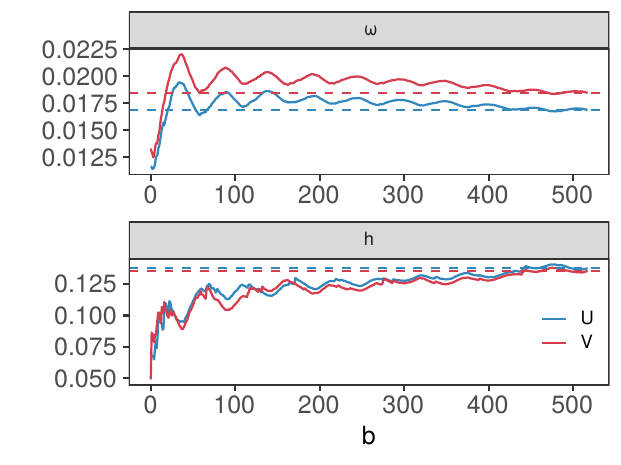}}\\
    \subfigure[$\mathrm{RFD}(\hat{\bm\omega}_{*}^{[b]},\hat{\bm\omega}_{*}^{\rm full})$]{
    \label{Fig:subfig:RFD_Omega_b}
    \includegraphics[scale=0.7]{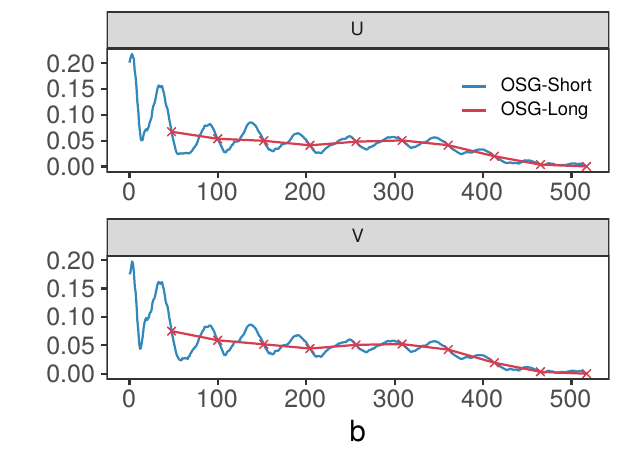}}
    \subfigure[$\mathrm{RFD}(\hat{\mathbf{h}}_{*}^{[b]},\hat{\mathbf{h}}_{*}^{\rm full})$]{
    \label{Fig:subfig:RFD_H_b}
    \includegraphics[scale=0.7]{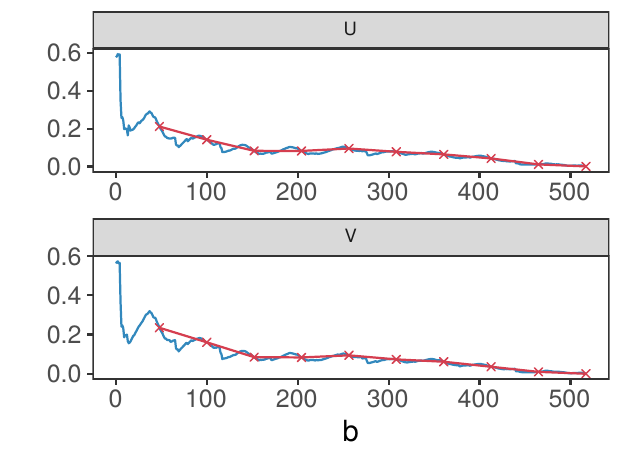}}
    \vspace{-10pt}
    \caption{Estimates and updates of parameters $\omega_{*,\alpha}$ and $h_{*,\alpha}$ in the Tukey h transformation for the second scenario. (a) compares the cumulative estimates up to the initial and final blocks. (b) illustrates the updates of $\omega_{*,1}^{[b]}$ and $h_{*,1}^{[b]}$ over successive blocks, with dashed lines indicating the estimates derived from the complete ten years of data. (c) displays the RFDs between the cumulative estimates of $\bm\omega_{*}=(\omega_{*,1},\ldots,\omega_{*,A})^\top$ and $\bm\omega_{*}^{\rm full}=(\omega_{*,1}^{\rm full},\ldots,\omega_{*,A}^{\rm full})^\top$. (d) shows the RFDs between the cumulative estimates of $\mathbf{h}_{*}=(h_{*,1},\ldots,h_{*,A})^\top$ and $\mathbf{h}_{*}^{\rm full}=(h_{*,1}^{\rm full},\ldots,h_{*,A}^{\rm full})^\top$. The $\times$ symbols in (c) and (d) mark the cumulative estimates up to each block in the first scenario.}
    \label{Fig:Converge_Tukeyh_Case2}
    \vspace{-20pt}
\end{figure}

Figs.~\ref{Fig:Converge_Tukeyh_Case2} and S6 demonstrate the estimates and updates of parameters $\omega_{*,\alpha}$ and $h_{*,\alpha}$ in the second scenario. As shown in Figs.~\ref{Fig:subfig:OmegaH_U_alpha_Case2} and S6(a), the significantly shorter initial block length results in substantial gaps between the initial estimates, $\hat{\omega}_{*,\alpha}^{[0]}$ and $\hat{h}_{*,\alpha}^{[0]}$, and the final estimates, $\hat{\omega}_{*,\alpha}^{[B]}$ and $\hat{h}_{*,\alpha}^{[B]}$. Observing the updates of $\hat{\omega}_{*,1}^{[b]}$ and $\hat{h}_{*,1}^{[b]}$ in Fig.~\ref{Fig:subfig:OmegaH_1_b_Case2}, we see that both parameters converge to the full dataset estimates as $b$ increases, though with initial large fluctuations that diminish over time. In general, $\hat{h}_{*,1}^{[b]}$ follows an upward trend, reflecting the arrival of new extreme values, while $\hat{\omega}_{*,1}^{[b]}$ trends downward. Notably,  $\hat{h}_{*,1}^{[b]}$ reaches $\hat{h}_{*,1}^{\rm full}$ around $b=460$ and shows small updates thereafter. This observation aligns with the findings from the first scenario, where $\hat{h}_{*,1}^{[b]}$ stabilizes after the ninth block. Both results imply that the data from the final year do not provide new information to further update the parameter $\hat{h}_{*,1}$. Fig.~\ref{Fig:Converge_Tukeyh_Case2}\hyperref[Fig:subfig:RFD_Omega_b]{(c)} and \hyperref[Fig:subfig:RFD_H_b]{(d)} display the evolutions of $\mathrm{RFD}(\hat{\bm\omega}_{*}^{[b]},\hat{\bm\omega}_{*}^{\rm full})$ and $\mathrm{RFD}(\hat{\mathbf{h}}_{*}^{[b]},\hat{\mathbf{h}}_{*}^{\rm full})$ over $b$ for both scenarios. In both scenarios, the RFDs decrease to zero, indicating convergence for all parameters. However, in the second scenario, the RFDs show more pronounced fluctuations, particularly when $b$ is small, suggesting less stability compared to the first scenario.

\begin{figure}[!b]
    \vspace{-10pt}
    \centering
    \subfigure[$\hat{\Phi}_1^{[b]}$ and $\hat{\Phi}_1^{\rm full}$]{
    \hspace{-10pt}
    \label{Fig:subfig:Phi1_Case2}
    \includegraphics[scale=0.48]{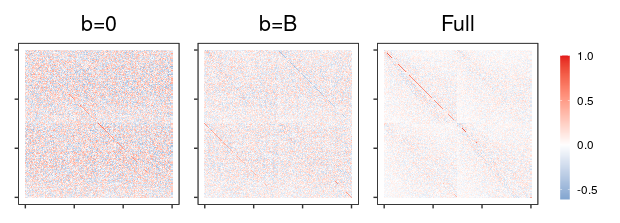}}
    \subfigure[$\hat{\mathbf{K}}^{[b]}$ and $\hat{\mathbf{K}}^{\rm full}$]{
    \label{Fig:subfig:K_Case2}
    \includegraphics[scale=0.48]{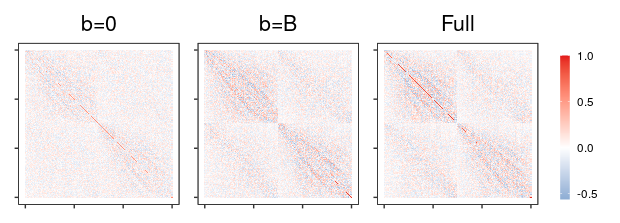}}\\
    \subfigure[$\mathrm{RDF}s$]{
    \hspace{-0.5in}
    \label{Fig:subfig:VAR2_Converge}
    \includegraphics[scale=0.7]{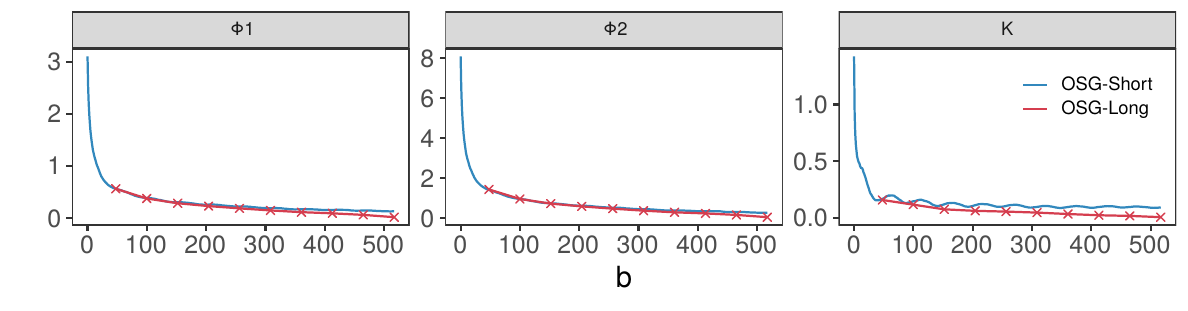}}
    \vspace{-10pt}
    \caption{Estimates and updates of the matrices $\Phi_1$ and $\mathbf{K}$ in the VAR($2$) model for the second scenario. (a) and (b) display the rescaled cumulative estimates of $\Phi$ and $\mathbf{K}$, respectively, up to the first block and the final block, along with their rescaled estimates obtained from the full ten-year data. All elements have been rescaled using the function $f(x)=\mathrm{sign}(x)\sqrt{|x|}$ for clearer visualization. (c) shows the RDFs between the cumulative estimates of $\Phi_1$, $\Phi_2$, and $\mathbf{K}$ and $\hat\Phi_1^{\rm full}$, $\hat\Phi_2^{\rm full}$, and $\hat{\mathbf{K}}^{\rm full}$, where $\times$ symbols mark the cumulative estimates up to each block in the first scenario.}
    \label{Fig:Converge_VAR2_Case2}
    \vspace{-20pt}
\end{figure}
Figs.~\ref{Fig:Converge_VAR2_Case2} and S7 illustrate estimates and updates of matrices $\Phi_1$, $\Phi_2$, and $\mathbf{K}$ for the second scenario. Due to the shorter initial block length, all estimates at $b=0$ fail to adequately capture the structures present in the full estimates, particularly for $\Phi_1$ and $\Phi_2$. In Fig.~\ref{Fig:subfig:VAR2_Converge}, there is a significant decrease in all RFDs before around $b=50$; however, as more blocks are added, the rate of decrease slows, which explains why the initial estimates in the first scenario resemble their full counterparts. Fluctuations in $\mathrm{RFD}(\hat{\mathbf{K}}^{[b]},\hat{\mathbf{K}}^{\rm full})$ indicate the instability of the updates. Furthermore, as $b$ increases, the loss of information in the second scenario becomes increasingly apparent.

\begin{figure}[!b]
\vspace{-10pt}
    \centering
    \subfigure[$\mathrm{I}_{\rm uq,U}$]{
    \label{Fig:subfig:Iuqu_Online_Case2}
    \includegraphics[scale=0.43]{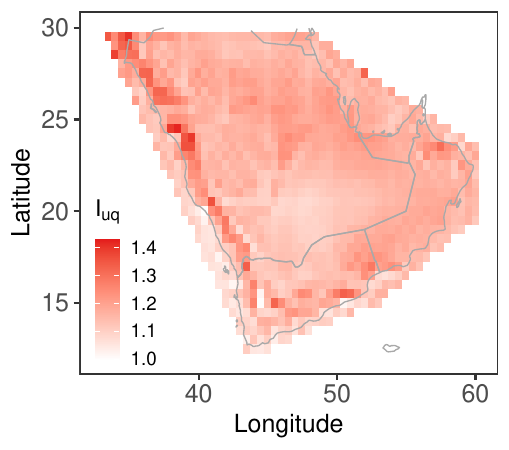}}
    \subfigure[$\mathrm{I}_{\rm wdt,U}$]{
    \label{Fig:subfig:Iwdtu_Online_Case2}
    \includegraphics[scale=0.43]{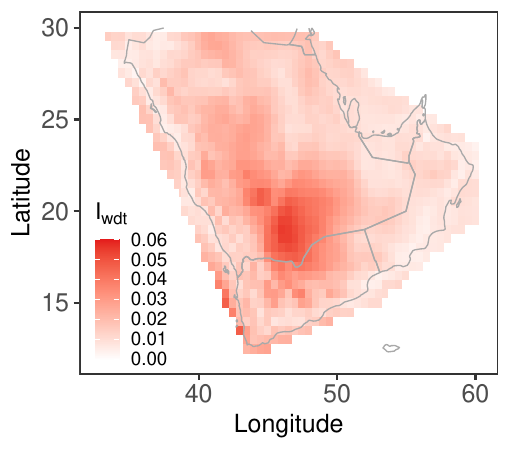}}
    \subfigure[$\mathrm{I}_{\rm wds,U}$]{
    \label{Fig:subfig:Iwdsu_Online_Case2}
    \includegraphics[scale=0.43]{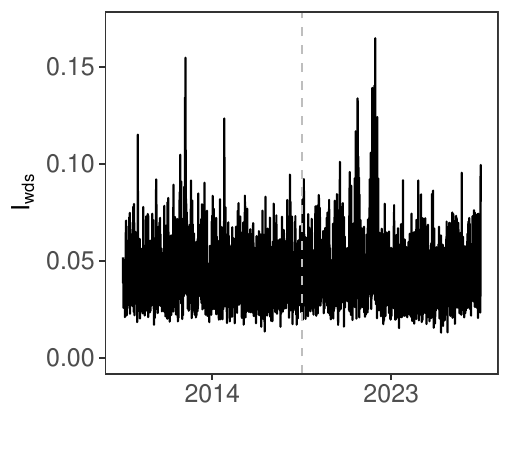}}
    \subfigure[$\mathrm{I}_{\rm rq,U}$]{
    \label{Fig:subfig:Irqu_Online_Case2}
    \includegraphics[scale=0.43]{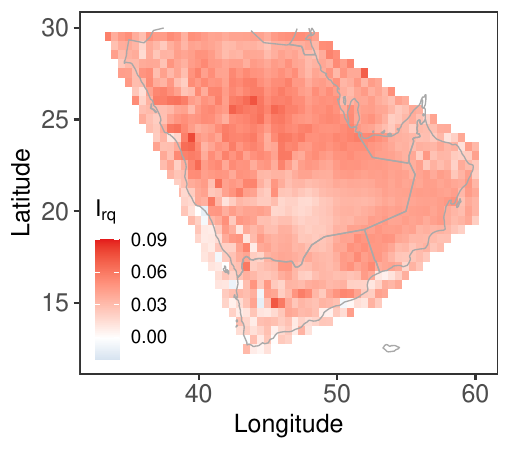}}\\
    \subfigure[$\mathrm{I}_{\rm uq,U}$]{
    \label{Fig:subfig:Boxplot_Iuqu_Case2}
    \includegraphics[scale=0.42]{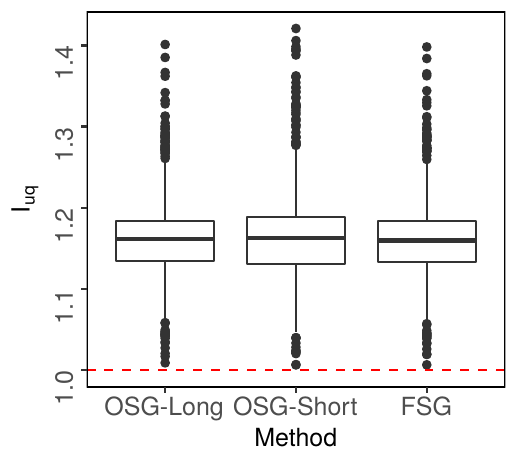}}
    \subfigure[$\mathrm{I}_{\rm wdt,U}$]{
    \label{Fig:subfig:Boxplot_Iwdtu_Case2}
    \includegraphics[scale=0.42]{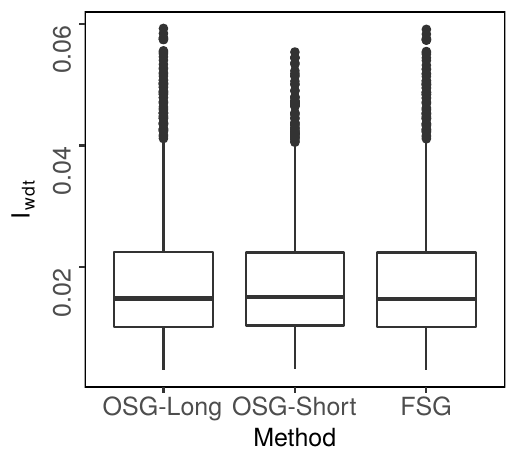}}
    \subfigure[$\mathrm{I}_{\rm wds,U}$]{
    \label{Fig:subfig:Boxplot_Iwdsu_Case2}
    \includegraphics[scale=0.42]{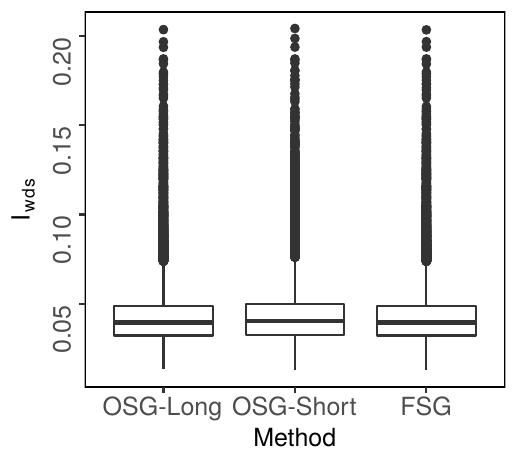}}
    \subfigure[$\mathrm{I}_{\rm rq,U}$]{
    \label{Fig:subfig:Boxplot_Irqu_Case2}
    \includegraphics[scale=0.42]{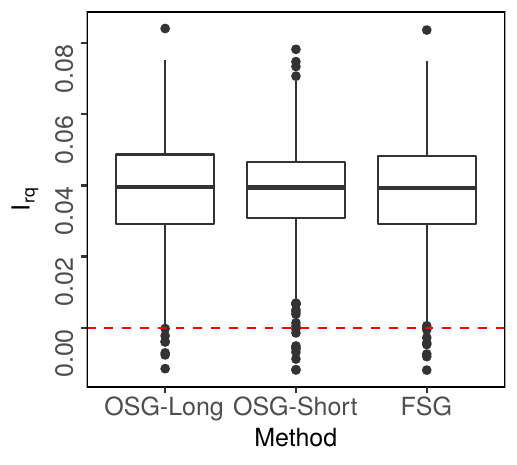}}\\
    \vspace{-10pt}
    \caption{Performance of the OSG constructed for the second scenario. (a)--(d) present various metrics used to assess the U-component emulations. (e)--(f) compare these metrics with those of the OSG-Long and the FSG. The red dashed lines in (e) and (h) mark the optimal index values.}
    \label{Fig:Assessment_Case2}
    \vspace{-20pt}
\end{figure}
\begin{figure}[!t]
    \vspace{-30pt}
    \centering
    \subfigure[ERA5 ensemble]{
    \label{Fig:subfig:Uncertaintyu_original}
    \includegraphics[scale=0.43]{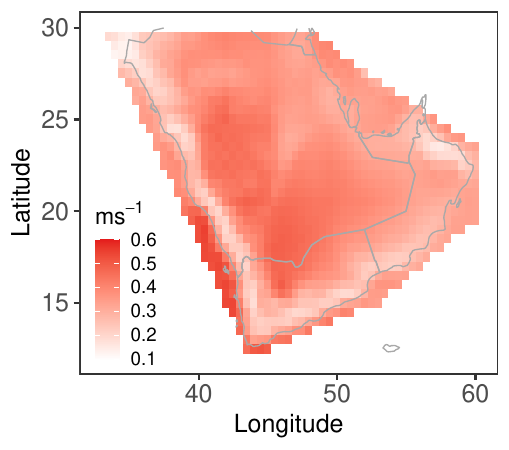}}
    \subfigure[FSG]{
    \label{Fig:subfig:Uncertaintyu_full}
    \includegraphics[scale=0.43]{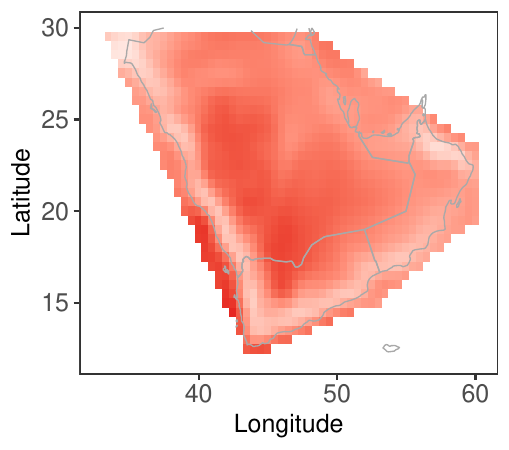}}
    \subfigure[OSG-Long]{
    \label{Fig:subfig:Uncertaintyu_Case1}
    \includegraphics[scale=0.43]{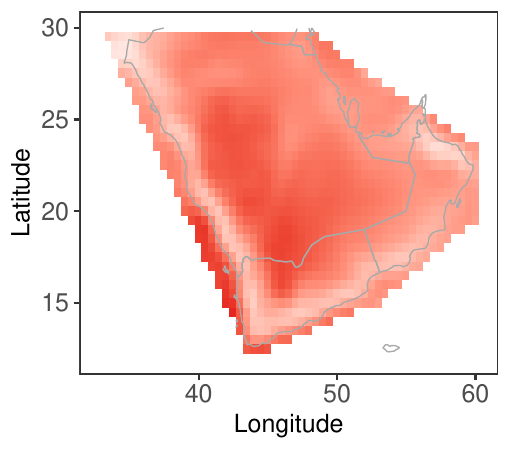}}
    \subfigure[OSG-Short]{
    \label{Fig:subfig:Uncertaintyu_Case2}
    \includegraphics[scale=0.43]{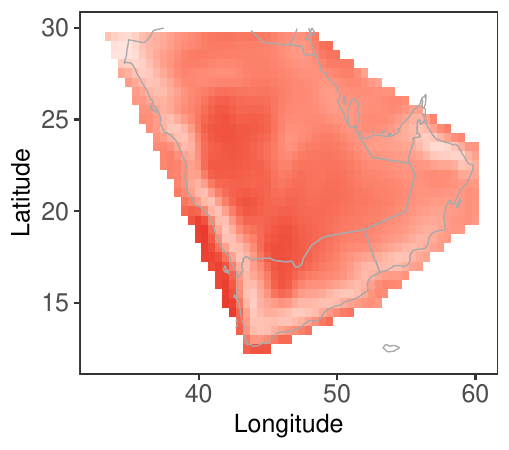}}\\
    \vspace{-10pt}
    \caption{Data unreliability over the ARP. (a) shows the map of $y_{U}^{\rm sd}(L_i,l_j)$, highlighting regions with relatively unreliable ERA5 ensembles. (b)--(d) present the maps of $\hat{y}_{U}^{\rm sd}(L_i,l_j)$.}
    \label{Fig:Uncertainty_Assessment}
\end{figure}

Figs.~\ref{Fig:Assessment_Case2} and S8 illustrate the performance of the OSG for the second scenario, showing that it performs well across all metrics. We further compare it with the OSG-Long and the FSG. The $\mathrm{I}_{\rm uq,U}$ index in Fig.~\ref{Fig:subfig:Iuqu_Online_Case2} exhibits higher values on the mountain ranges compared to Fig.~\ref{Fig:subfig:Iuqu_Online_Case1}, indicating a slight loss in accuracy in capturing the dependence structures. The $\mathrm{I}_{\rm wdt,U}$ values over the central plateau region in Fig.~\ref{Fig:subfig:Iwdtu_Online_Case2} are slightly lower than those in Fig.~\ref{Fig:subfig:Iwdtu_Online_Case1}. However, as shown in Fig.~\ref{Fig:subfig:Boxplot_Iwdtu_Case2}, the median $\mathrm{I}_{\rm wdt,U}$ of the fast-update OSG-Short ($0.0150$) is slightly higher than those of the OSG-Long ($0.0148$) and the FSG ($0.0148$). All SGs have comparable $\mathrm{I}_{\rm wds,U}$ values. In Figs.~\ref{Fig:subfig:Irqu_Online_Case2} and \ref{Fig:subfig:Irqu_Online_Case1}, the maps of $\mathrm{I}_{\rm rq,U}$ exhibit distinct patterns, meaning that regions where data unreliability is overestimated differ between cases. However, Figs.~\ref{Fig:Uncertainty_Assessment} and S9 show that all emulations are capable of identifying regions with higher data unreliability.

\section{Conclusion and Extensions}
\label{sec:Discussion}
This paper proposes an OSG algorithm for bivariate wind speed ERA5 ensembles over any specific global region and constructs OSG-Long and OSG-short for two different scenarios. These OSGs generate fast stochastic approximations for ERA5 ensembles of size $2RT|\mathcal{G}_{\mathcal{R}}|$ while requiring only $4T|\mathcal{G}_{\mathcal{R}}|+4A+4A^2P+4A^2$ stored parameters. Compared to existing SGs, the OSGs significantly reduce storage needs during development and enable near real-time emulations. They allow data to be sequentially incorporated into the model in either long or short blocks, with each block updating the model parameters. Consequently, the cumulative parameter estimates converge to those derived from the full dataset, making the OSGs comparable in performance to the FSG.

When designing the OSG algorithm, we begin with modeling a single data block. First, we employ a spatial mixed effects model to identify and remove fixed effects from the data. Then, we use Slepian bases, which are optimally concentrated on the given region of interest and independent from the data, to represent the data in a lower-dimensional space and obtain Slepian coefficients. Following this, we apply the Tukey h transformation and the Lambert W function to Gaussianize the Slepian coefficients. Finally, we implement a VAR($2$) model to capture the dependence structure of the transformed coefficients. The procedures for building an SG for a single block are summarized in  Algorithm~S1, while Algorithm~S2 outlines the process for generating emulations. We then derive the update formulas for all model parameters, which combine cumulative estimates with information from the current block. The updating procedures are summarized in Algorithm~S3. A key feature of our approach is the application of Slepian concentration technique. The optimal concentration of Slepian bases ensures efficient representation of the data using the fewest bases necessary. Their data independence ensures their efficiency across all variables and blocks. Furthermore, Lambert W functions play a crucial role in providing closed-form estimates for the parameters in the Tukey h transformation.

The proposed OSG offers significant potential for various extensions. For instance, the reduction in storage requirements facilitate the emulation of climate data with higher spatial and temporal resolutions, as well as a broader range of variables. Moreover, the construction of the OSG could be expanded beyond a single region to cover multiple regions of interest or even the entire globe. Existing global SGs often have to balance across diverse regions, as variables may exhibit distinct characteristics in different areas, leading to the loss of critical information in joint analyses. By dividing the globe into regions based on specific criteria and developing OSGs for each region in parallel, a more accurate and powerful global SG can be achieved.

\vspace{5pt}
\hspace{-18pt}\textbf{Supplementary Materials:} Additional information about the ERA5 data, algorithms, and results is provided in the supplementary materials. (.pdf file)
\vspace{-20pt} 


\baselineskip 19.5pt
\bibliographystyle{asa}
\bibliography{Ref}

\newpage
\smallskip
\begin{center}
{\large\bf Supplementary Materials for ``Online stochastic generators using Slepian bases for regional bivariate wind speed ensembles from ERA5"}
\end{center}

\setcounter{lemma}{0}  
\renewcommand{\thelemma}{S\arabic{lemma}}
\setcounter{section}{0}  
\renewcommand{\thesection}{S\arabic{section}}
\setcounter{algorithm}{0}  
\renewcommand{\thealgorithm}{S\arabic{algorithm}}
\setcounter{equation}{0} 
\renewcommand{\theequation}{S\arabic{equation}}
\setcounter{figure}{0} 
\renewcommand{\thefigure}{S\arabic{figure}}
\setcounter{table}{0} 
\renewcommand{\thetable}{S\arabic{table}}
\setcounter{page}{1} 
\renewcommand{\thepage}{S\arabic{page}}

\section{Introduction}
This document supplements the main manuscript, providing more details about the Arabian-Peninsula region (ARP), methodology, validation, and results in case studies.

\section{Supplement to Data Description}
\label{sec:Details_of_ARP}
\begin{figure}[!b]
    \centering
    \includegraphics[scale=0.4]{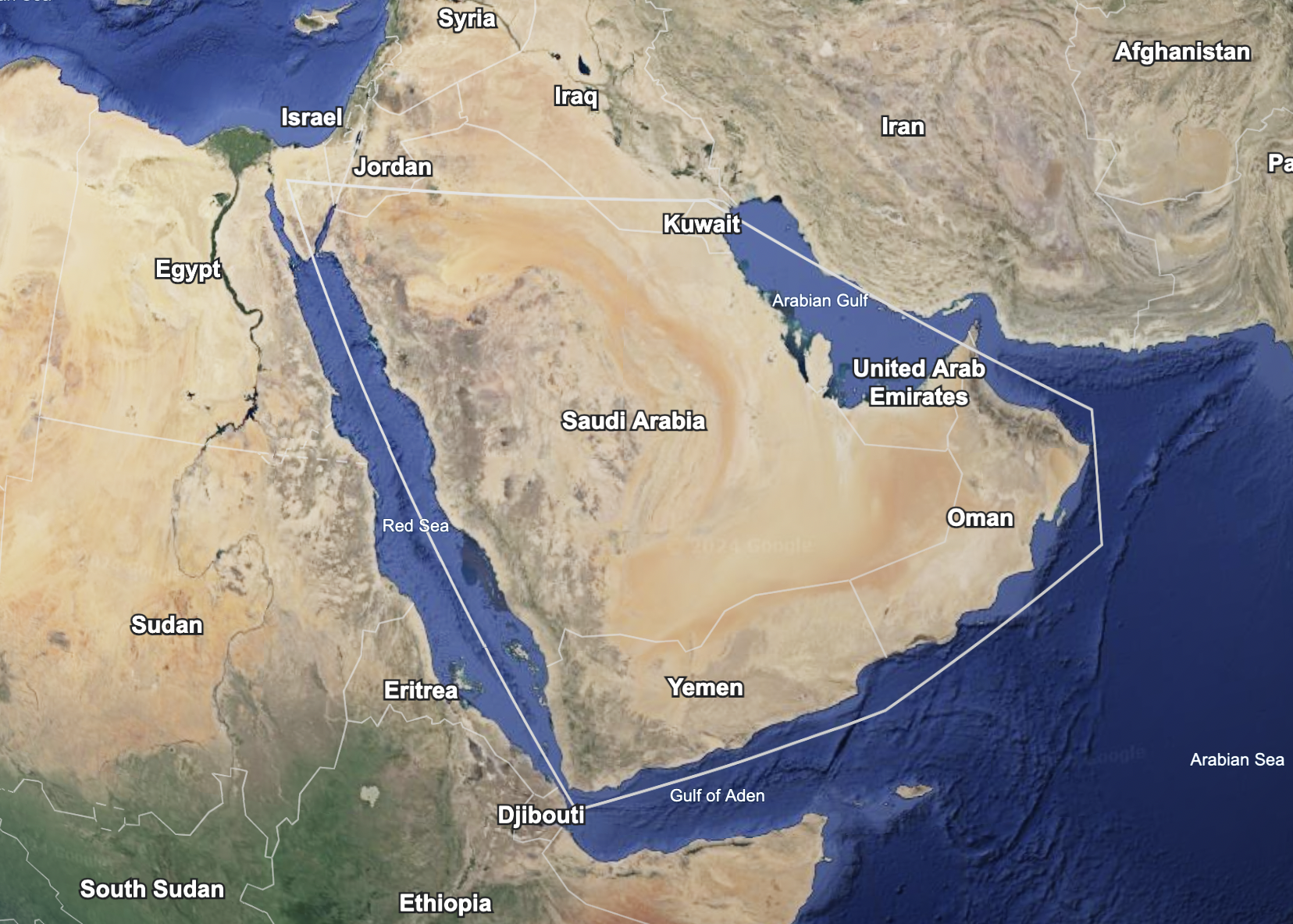}
    \caption{Topography of the ARP and its surrounding area depicted using Google Earth imagery, with the ARP delineated by solid white segments.}
    \label{fig:ARP_MAP}
\end{figure}
As shown in Fig.~\ref{fig:ARP_MAP}, the ARP defined by \citet{essd-12-2959-2020} is surrounded by the Red Sea to the west, the Arabian Sea to the southeast, and the Gulf of Oman and the Arabian Gulf to the east. Along the western edge of the ARP, several mountain ranges run parallel to the Red Sea. Additionally, there are mountain ranges stretch along the northeastern coast of Oman and the eastern United Arab Emirates. These mountains rise sharply from the coastal plains and can reach heights exceeding 3,000 meters in some areas, significantly influencing local wind patterns. Moreover, the ARP has plateaus in its interior regions. Deserts, depicted in darker sand colors, cover the northern, central, and southern regions of the ARP.

Located within the trade-wind belt of the Northern Hemisphere, ARP experiences predominant northeasterly trade winds, resulting in a general downward shift across all curves in Fig.~1(c). In winter and early spring, the Shamal wind from the west brings eastward and southward flows. During summer, the Indian monsoon introduces southeasterly winds from the Indian Ocean, particularly affecting the southeastern ARP. Elsewhere in ARP, westerly or northwesterly winds may prevail. Autumn marks a transitional period with weakening winds.

\section{Supplement to Methodology}
\subsection{Algorithms}
Algorithms~\ref{alg:SingleSG_Model} and \ref{alg:SingleSG_Emulation} outline the procedures of developing an SG and generating emulations for regional bivariate wind speed ERA5 ensembles over a fixed time period, respectively. By setting $\mathcal{T}_0=\{1,\ldots,T\}$, we can develop the SG and generate emulations for the entire dataset. Based on Algorithm~\ref{alg:SingleSG_Model} and the online updating formulas, Algorithm~\ref{alg:OnlineSG_Model} summarizes the steps for constructing an OSG. Using the outputs from Algorithm~\ref{alg:OnlineSG_Model}, we can generate emulations for the entire dataset with Algorithm~\ref{alg:SingleSG_Emulation}.

\begin{algorithm}[!b]
\caption{Construct an SG for wind speed ERA5 ensembles over a fixed period}
\label{alg:SingleSG_Model}
\begin{algorithmic}[3] 
\renewcommand{\algorithmicrequire}{ \textbf{Input:}} 
\REQUIRE $A$, $P$, $\{y_{*,t}^{(r)}(L_i,l_j)\}_{t\in\mathcal{T}_0, r\in\Upsilon, (L_i,l_j)\in\mathcal{G}_{\mathcal{R}}}$.\\
\vspace{8pt}
(1) For each $t\in\mathcal{T}_0$ and $(L_i,l_j)\in\mathcal{G}_{\mathcal{R}}$, evaluate and remove $m_{*,t}(L_i,l_j)$ from \\
\hspace{5mm} $\{y_{*,t}^{(r)}(L_i,l_j)\}_{r\in\Upsilon}$ to obtain the random effects $\{z_{*,t}^{(r)}(L_i,l_j)\}_{r\in\Upsilon}$.\\
\vspace{8pt}
(2) For each $t\in\mathcal{T}_0$ and $r\in\Upsilon$, apply Slepian concentration to $\{z_{*,t}^{(r)}(L_i,l_j)\}_{(L_i,l_j)\in\mathcal{G}_{\mathcal{R}}}$ to \\
\hspace{5mm} obtain the Slepian coefficients $\{s_{*,t}^{(r)}(\alpha)\}_{\alpha\in\mathcal{A}}$, where $\mathcal{A}=\{1,\ldots,A\}$, and then eval-\\
\hspace{5mm} uate $\{\sigma_{*,t}(l_i,l_j)\}_{(L_i,l_j)\in\mathcal{G}_{\mathcal{R}}}$. \\
\vspace{8pt}
(3) For each $\alpha\in\mathcal{A}$, evaluate $\gamma_{*,\alpha}$ and $\kappa_{*,\alpha}$ using (8), $\omega_{*,\alpha}$ and $h_{*,\alpha}$ using (7), and \\
\hspace{5mm} then perform the Tukey h transformation in (5) on $\{s_{*,t}^{(r)}(\alpha)\}_{t\in\mathcal{T}_0, r\in\Upsilon}$ to obtain the \\ \hspace{5mm} transformed coefficients $\{\tilde{s}_{*,t}^{(r)}(\alpha)\}_{t\in\mathcal{T}_0, r\in\Upsilon}$.\\
\vspace{8pt}
(4) Evaluate matrices $\Phi$ and $\mathbf{K}$ in the VAR(P) model using (10) and (11), respectively.\\
\vspace{8pt}
\renewcommand{\algorithmicensure}{ \textbf{Output:}} 
\ENSURE $\{\hat{m}_{*,t}(L_i,l_j),\hat{\sigma}_{*,t}(L_i,l_j)\}_{t\in\mathcal{T}_0, (L_i,l_j)\in\mathcal{G}_{\mathcal{R}}}$, $\{\hat{\gamma}_{*,\alpha},\hat{\kappa}_{*,\alpha}\}_{\alpha\in\mathcal{A}}$, $\hat{\Phi}$, and $\hat{\mathbf{K}}$.\\ 
\end{algorithmic}
\end{algorithm}

\begin{algorithm}[!t]
\caption{Emulate wind speed ERA5 ensembles over a fixed period}
\label{alg:SingleSG_Emulation}
\begin{algorithmic}[3] 
\renewcommand{\algorithmicrequire}{ \textbf{Input:}} 
\REQUIRE $\{\hat{m}_{*,t}(L_i,l_j),\hat{\sigma}_{*,t}(L_i,l_j)\}_{t\in\mathcal{T}_0, (L_i,l_j)\in\mathcal{G}_{\mathcal{R}}}$, $\{\hat{\gamma}_{*,\alpha},\hat{\kappa}_{*,\alpha}\}_{\alpha\in\mathcal{A}}$, $\hat{\Phi}$ and $\hat{\mathbf{K}}$, $R'$.\\
\vspace{8pt}
For $r=1,\ldots,R'$,\\
\vspace{8pt}
(1) for $t=1,\ldots,|\mathcal{T}_0|$, generate $\bm\xi_{t}^{(r)}\sim\mathcal{N}_{2|\mathcal{A}|}(\mathbf{0},\hat{\mathbf{K}})$, then use it along with $\hat\Phi$ and (9) to \\
\hspace{5mm} calculate the coefficient vector $\tilde{\mathbf{s}}_t^{(r)}$, where the required $\tilde{\mathbf{s}}_{-(p-1)}^{(r)}$, $p=1,\ldots,P$, are\\
\hspace{5mm} also generated from $\mathcal{N}_{2|\mathcal{A}|}(\mathbf{0},\hat{\mathbf{K}})$,\\
\vspace{8pt}
(2) for each $\alpha\in\mathcal{A}$, calculate $\hat\omega_{*,\alpha}$ and $\hat{h}_{*,\alpha}$ using $\hat{\gamma}_{*,\alpha}$, $\hat{\kappa}_{*,\alpha}$ and (7), and then perform\\
\hspace{5mm} the inverse Tukey h transformation in (6) on $\{\tilde{s}_{*,t}^{(r)}(\alpha)\}_{t\in\mathcal{T}_0}$ to obtain the Slepian\\
\hspace{5mm} coefficients $\{s_{*,t}^{(r)}(\alpha)\}_{t\in\mathcal{T}_0}$,\\
\vspace{8pt}
(3) for each $t\in\mathcal{T}_0$, apply the inverse Slepian concentration to $\{s_{*,t}^{(r)}(\alpha)\}_{\alpha\in\mathcal{A}}$ to obtain\\
\hspace{5mm} $\sum_{\alpha\in\mathcal{A}}s_{*,t}^{(r)}(\alpha)g_{\alpha}(L_i,l_j)$ in (1),\\
\vspace{8pt}
(4) for each $t\in\mathcal{T}_0$ and $(L_i,l_j)\in\mathcal{G}_{\mathcal{R}}$, 
generate the residual $\epsilon_{*,t}^{(r)}(L_i,l_j)\sim\mathcal{N}(0,\hat{\sigma}^2(L_i,l_j))$, \\
\hspace{5mm} add $\sum_{\alpha\in\mathcal{A}}s_{*,t}^{(r)}(\alpha)g_{\alpha}(L_i,l_j)$ to obtain the random effect $z_{*,t}^{(r)}(L_i,l_j)$, and then add the\\
\hspace{5mm} ensemble mean $\hat{m}_{*,t}(L_i,l_j)$ to get the emulation $\hat{y}_{*,t}^{(r)}(L_i,l_j)$.  \\
\vspace{8pt}
\renewcommand{\algorithmicensure}{ \textbf{Output:}} 
\ENSURE $\{\hat{y}_{*,t}^{(r)}(L_i,l_j)\}_{t\in\mathcal{T}_0, (L_i,l_j)\in\mathcal{G}_{\mathcal{R}}}$.\\ 
\end{algorithmic}
\end{algorithm}

\vspace{50pt}

\begin{algorithm}[!h]
\caption{Construct an OSG for regional bivariate wind speeds ERA5 ensembles}
\label{alg:OnlineSG_Model}
\begin{algorithmic}[3] 
\renewcommand{\algorithmicrequire}{ \textbf{Input:}} 
\REQUIRE $A$, $P$, $\{y_{*,t}^{(r)}(L_i,l_j)\}_{t\in\mathcal{T}_0, r\in\Upsilon, (L_i,l_j)\in\mathcal{G}_{\mathcal{R}}},\ldots,\{y_{*,t}^{(r)}(L_i,l_j)\}_{t\in\mathcal{T}_B, r\in\Upsilon, (L_i,l_j)\in\mathcal{G}_{\mathcal{R}}}$.\\
\vspace{8pt}
(1) For the initial data block $\{y_{*,t}^{(r)}(L_i,l_j)\}_{t\in\mathcal{T}_0, r\in\Upsilon, (L_i,l_j)\in\mathcal{G}_{\mathcal{R}}}$, follow Algorithm~\ref{alg:SingleSG_Model} to \\
\hspace{5mm} obtain $\{\hat{m}_{*,t}(L_i,l_j),\hat{\sigma}_{*,t}(L_i,l_j)\}_{t\in\mathcal{T}_0, (L_i,l_j)\in\mathcal{G}_{\mathcal{R}}}$, $\{\hat{\gamma}_{*,\alpha}^{[0]},\hat{\kappa}_{*,\alpha}^{[0]}\}_{\alpha\in\mathcal{A}}$, $\hat{\Phi}^{[0]}$ and $\hat{\mathbf{K}}^{[0]}$.\\
\hspace{5mm} Additionally, store $\mathbf{X}^{[0]}$ from step (4) of Algorithm~\ref{alg:SingleSG_Model}.\\
\vspace{8pt}
(2) For the $(b+1)$th data block $\{y_{*,t}^{(r)}(L_i,l_j)\}_{t\in\mathcal{T}_b, r\in\Upsilon, (L_i,l_j)\in\mathcal{G}_{\mathcal{R}}}$, $b=1,\ldots,B$,\\
\vspace{8pt}
\hspace{5mm} (a) follow steps (1) and (2) of Algorithm~\ref{alg:SingleSG_Model} to obtain the block of Slepian coefficients \\
\hspace{10mm} $\{s_{*,t}^{(r)}(\alpha)\}_{t\in\mathcal{T}_b, r\in\Upsilon, \alpha\in\mathcal{A}}$ and $\{\hat{m}_{*,t}(L_i,l_j),\hat{\sigma}_{*,t}(L_i,l_j)\}_{t\in\mathcal{T}_b, (L_i,l_j)\in\mathcal{G}_{\mathcal{R}}}$.\\
\vspace{8pt}
\hspace{5mm} (b) follow step (3) of Algorithm~\ref{alg:SingleSG_Model} to obtain the block of transformed coefficients \\
\hspace{10mm} $\{\tilde{s}_{*,t}^{(r)}(\alpha)\}_{t\in\mathcal{T}_b, r\in\Upsilon, \alpha\in\mathcal{A}}$ and $\{\hat{\gamma}_{*,\alpha}^{\{b\}},\hat{\kappa}_{*,\alpha}^{\{b\}}\}_{\alpha\in\mathcal{A}}$, then use (12) and (13) to update\\
\hspace{10mm} $\{\hat{\gamma}_{*,\alpha}^{[b-1]},\hat{\kappa}_{*,\alpha}^{[b-1]}\}_{\alpha\in\mathcal{A}}$ to be $\{\hat{\gamma}_{*,\alpha}^{[b]},\hat{\kappa}_{*,\alpha}^{[b]}\}_{\alpha\in\mathcal{A}}$.\\
\vspace{8pt}
\hspace{5mm} (c) follow step (4) of Algorithm~\ref{alg:SingleSG_Model} to obtain $\mathbf{X}^{\{b\}}$, $\hat{\Phi}^{\{b\}}$ and $\hat{\mathbf{K}}^{\{b\}}$, then use (14) and\\
\hspace{10mm} (15) to update $\hat{\Phi}^{[b-1]}$ and $\hat{\mathbf{K}}^{[b-1]}$ to be $\hat{\Phi}^{[b]}$ and $\hat{\mathbf{K}}^{[b]}$, respectively.\\
\vspace{7pt}
\renewcommand{\algorithmicensure}{ \textbf{Output:}} 
\ENSURE $\{\hat{m}_{*,t}(L_i,l_j),\hat{\sigma}_{*,t}(L_i,l_j)\}_{t\in\cup_{b=0}^B\mathcal{T}_b, (L_i,l_j)\in\mathcal{G}_{\mathcal{R}}}$, $\{\hat{\gamma}_{*,\alpha}^{[B]},\hat{\kappa}_{*,\alpha}^{[B]}\}_{\alpha\in\mathcal{A}}$, $\mathbf{X}^{[B]}$, $\hat{\Phi}^{[B]}$ and $\hat{\mathbf{K}}^{[B]}$.\\ 
\end{algorithmic}
\end{algorithm}

\subsection{Online parameter updating for the Tukey g-and-h transformation}
This subsection presents a sequential strategy for updating parameters in the Tukey g-and-h (TGH) transformation. We begin by assuming that the data $\{s_t\}_{t\in\mathcal{T}}$ is only skewed and satisfies the Tukey g transformation:
\begin{equation*}
    s_t=g^{-1}\{\exp(g\Tilde{s}_t)-1\},
\end{equation*}
where $\mathcal{T}=\{1,\ldots,T\}$ and $\Tilde{s}_t\sim\mathcal{N}(0,1)$. Then, estimating $g$ can be done by maximizing the log-likelihood function
\begin{equation*}
    l(g)=\sum_{t=1}^Tl_t(g)=-\frac{T}{2}\log(2\pi)-\frac{1}{2}\sum_{t=1}^T\left\{\frac{\log(gs_i+1)}{g}\right\}^2,
\end{equation*}
or equivalently, solving an estimating equation (EE)
\begin{equation*}
    \sum_{t=1}^T\frac{\partial l_t(g)}{\partial g}=\sum_{t=1}^T\left[\frac{\{\log(gs_i+1)\}^2}{g^3}-\frac{s_i\log(gs_i+1)}{g^2(gs_i+1)}\right]=0,
\end{equation*}
or equivalently, solving another EE
\begin{equation*}
    \Psi_{\mathcal{T}}(g)=\sum_{t=1}^T\psi_t(g)=\sum_{t=1}^T[(gs_i+1)\{\log(gs_i+1)\}^2-gs_i\log(gs_i+1)]=0.
\end{equation*}
Denote $\hat g_{\mathcal{T}}$ to be the solution of the above EE and $g_0$ is the solution of $\mathrm{E}\{\psi(g)\}=0$. Similarly, for each data block $\mathcal{S}^{\{b\}}$, we have $\hat{g}^{\{b\}}$ be the solution of
\begin{equation*}
    \Psi_{\mathcal{T}_b}(g)=\sum_{t\in\mathcal{T}_b}\psi_t(g)=0.
\end{equation*}
\vspace{-10pt}
The Taylor expansion of $\Psi_{\mathcal{T}_b}(g)$ at $\hat g^{\{b\}}$ is
\begin{equation}
\begin{aligned}
    \Psi_{\mathcal{T}_b}(g)&=\Psi_{\mathcal{T}_b}(\hat g^{\{b\}})+\sum_{t\in\mathcal{T}_b}\frac{\partial \psi_t(g)}{\partial g}|_{g=\hat g^{\{b\}}}(g-\hat g^{\{b\}})+r\\
    &\approx\beta^{\{b\}}(g-\hat g^{\{b\}}),
\end{aligned}
\label{eq:Psi_b}
\end{equation}
where the second equality is derived from facts that $\Psi_{\mathcal{T}_b}(\hat g^{\{b\}})=0$ and $r$ is negligible. Here 
\begin{equation*}
    \beta^{\{b\}}=\sum_{t\in\mathcal{T}_b}\frac{\partial \psi_t(g)}{\partial g}|_{g=\hat g^{\{b\}}}
\end{equation*}
only relies on the $(b+1)$th block. From~\eqref{eq:Psi_b}, we let the cumulative estimator $\hat g^{[B]}$ be the solution
\begin{equation*}
    \Psi_{\mathcal{T}}(g)=\sum_{b=0}^B\Psi_{\mathcal{T}_b}(g)=\sum_{b=0}^B\beta^{\{b\}}(g-\hat{g}^{\{b\}})=0.
\end{equation*}
That is, 
\begin{equation*}
    \hat g^{[B]}=\left(\sum_{b=0}^B\beta^{\{b\}}\right)^{-1}\sum_{b=0}^B\beta^{\{b\}}\hat{g}^{\{b\}}.
\end{equation*}
Therefore, the update of $\hat g^{[b]}$ using $\hat g^{\{b\}}$ and $\hat g^{[b-1]}$ can be done by
\begin{equation*}
\begin{aligned}
    \hat g^{[b]}&=\left(\sum_{\iota=0}^{b-1}\beta^{\{\iota\}}+\beta^{\{b\}}\right)^{-1}\left(\sum_{\iota=0}^{b-1}\beta^{\{\iota\}} \hat{g}^{[b-1]}+\beta^{\{b\}} \hat{g}^{\{b\}}\right)\\
    &=(\beta^{[b-1]}+\beta^{\{b\}})^{-1}(\beta^{[b-1]}\hat{g}^{[b-1]}+\beta^{b}\hat{g}^{\{b\}}).
\end{aligned}
\end{equation*}
Therefore, except for storing $\hat g^{[b]}$, we need to store $\beta^{[b-1]}$.

Now, we assume that the data $\{s_t\}_{t=1}^T$ satisfies a TGH:
\begin{equation*}
    s_t=\omega g^{-1}\{\exp(g\Tilde{s}_t)-1\}\exp\{h\Tilde{s}_t^2/2\}.
\end{equation*}
We outline a sequential procedure to update the parameters. Given the initial data block, we can obtain $\hat{\omega}^{[0]}$, $\hat{g}^{[0]}$, and $\hat{h}^{[0]}$ by existing methods using quantiles or approximated likelihoods. For $b=1,\ldots,B$, we first calculate $\hat{\omega}^{\{b\}}$, $\hat{g}^{\{b\}}$, and $\hat{h}^{\{b\}}$ based on the current data block, and then do the inverse TGH to get $\{\Tilde{s}_t\}_{t\in\mathcal{T}_b}$. After that, we plug $\hat{h}^{[b-1]}$ and $\hat{\omega}^{[b-1]}$ into the TGH, and use the above method to update $\hat{g}^{[b-1]}$ to be $\hat{g}^{[b]}$. Finally, we plug $\hat{g}^{[b]}$ into TGH and update $\hat h$ and $\hat\omega$.

\section{Supplementary to Case Studies}
\subsection{Supplement to parameter selection}
\label{sec:subsec:Supplement_to_Parameters_Selection}
\begin{figure}[!t]
    \centering
    \subfigure[$\mathrm{I}_{\rm uq,*}$]{
    \label{Fig:subfig:ChooseA_Iuq}
    \includegraphics[scale=0.49]{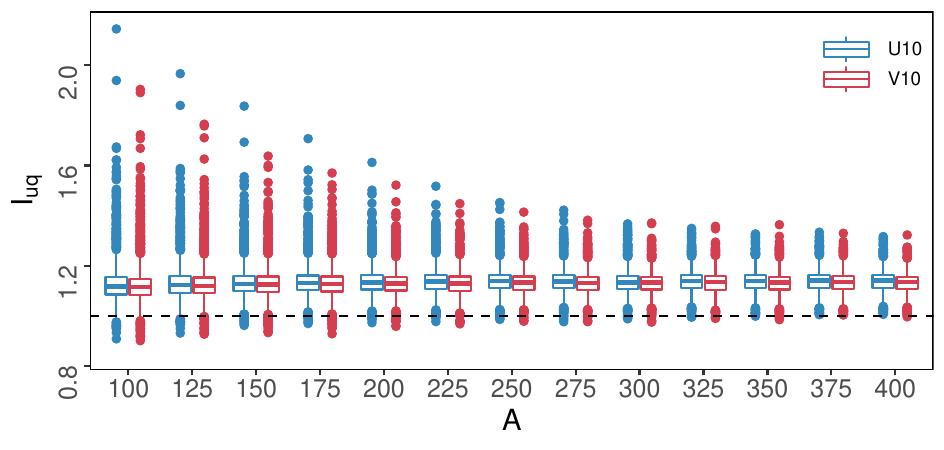}}
    \subfigure[$\mathrm{I}_{\rm wdt,*}$]{
    \label{Fig:subfig:ChooseA_Iwdt}
    \includegraphics[scale=0.49]{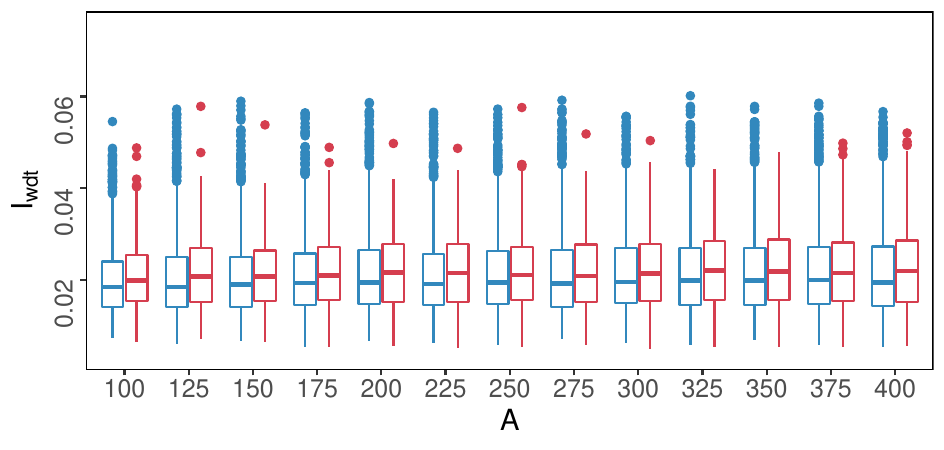}}\\
    \subfigure[$\mathrm{I}_{\rm wds,*}$]{
    \label{Fig:subfig:ChooseA_Iwds}
    \includegraphics[scale=0.49]{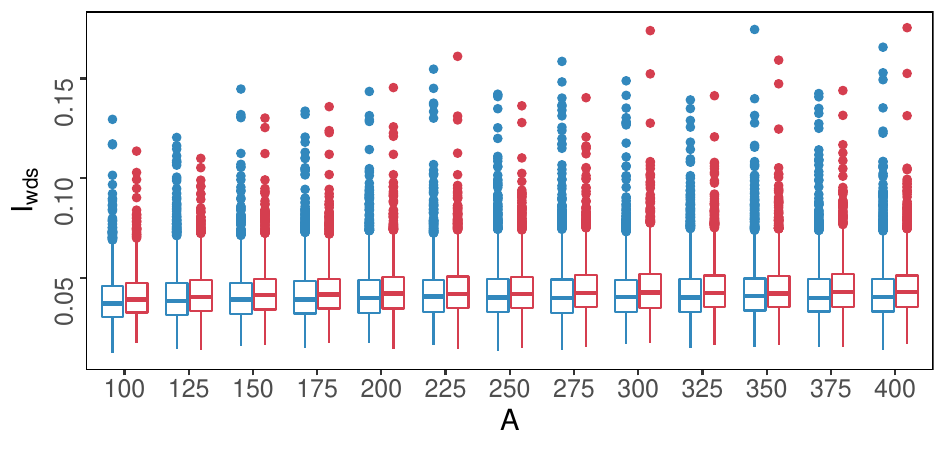}}
    \subfigure[$\mathrm{I}_{\rm rq,*}$]{
    \label{Fig:subfig:ChooseA_Irq}
    \includegraphics[scale=0.49]{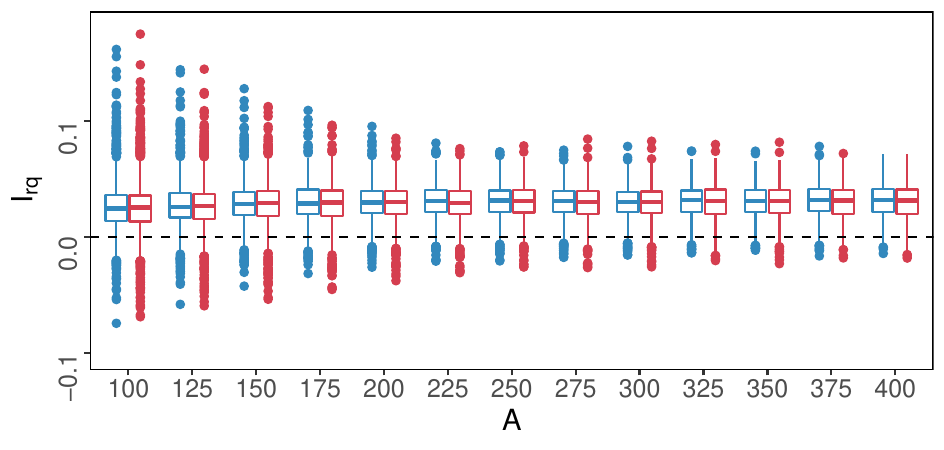}}\\
    \subfigure[$\mathrm{I}_{\rm uq}$]{
    \label{Fig:subfig:ChooseA_Iuq_Combine}
    \includegraphics[scale=0.49]{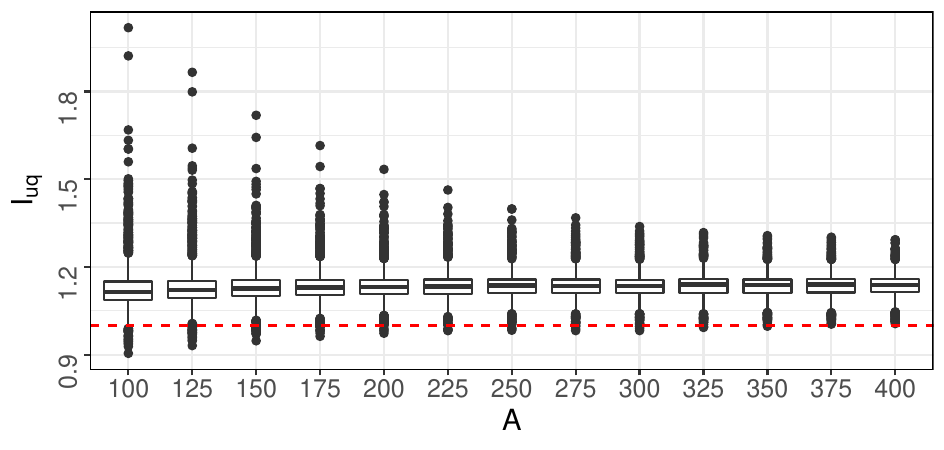}}
    \subfigure[$\mathrm{I}_{\rm rq}$]{
    \label{Fig:subfig:ChooseA_Irq_Combine}
    \includegraphics[scale=0.49]{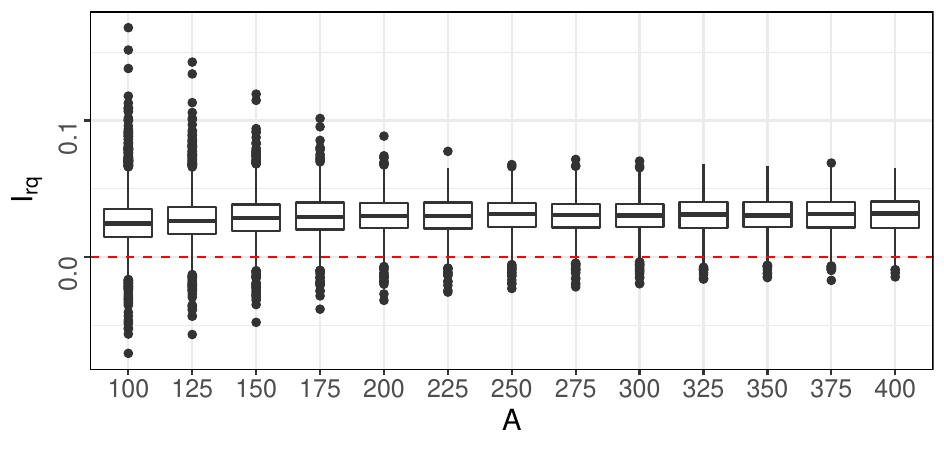}}\\
    \subfigure[$\mathrm{I}_{\rm uq,U}$]{
    \label{Fig:subfig:Iuq_A100_A300}
    \includegraphics[scale=0.49]{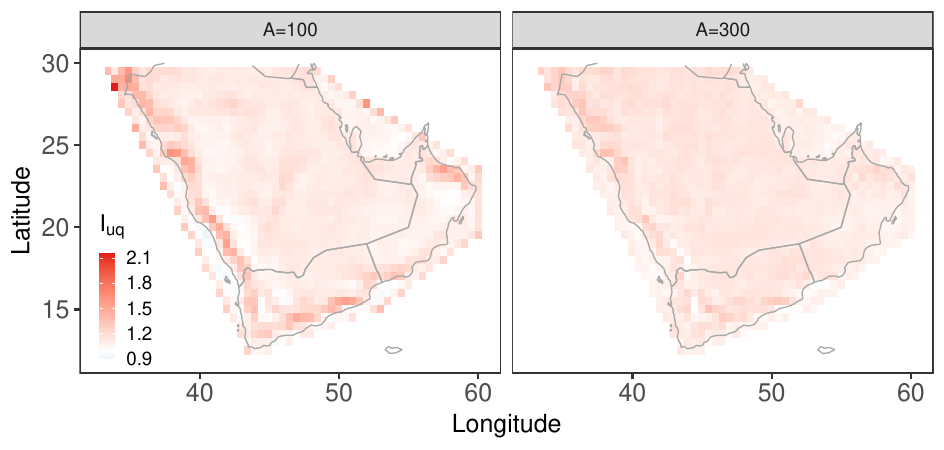}}
    \subfigure[$\mathrm{I}_{\rm wdt,U}$]{
    \label{Fig:subfig:Iwdt_A100_A300}
    \includegraphics[scale=0.49]{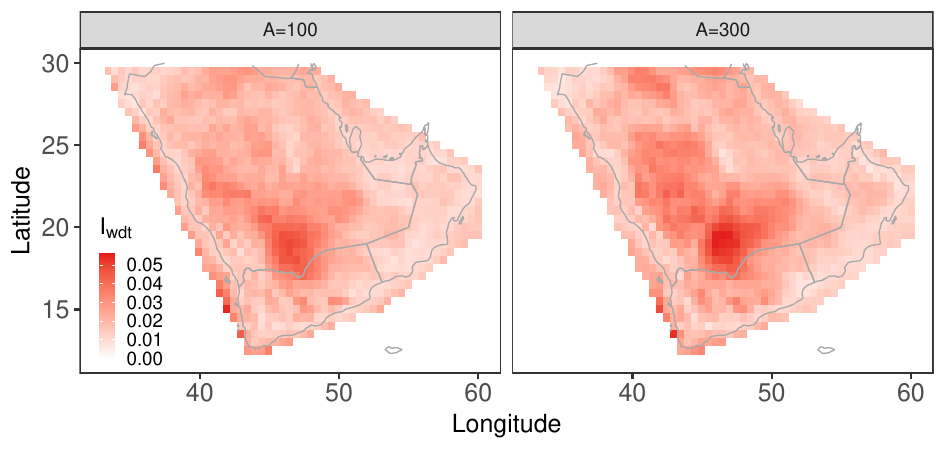}}\\ 
    \caption{Selection of $A$ in the Slepian concentration. (a)--(d) present boxplots of the indices $\{\mathrm{I}_{\rm uq,*}(L_i,l_j)\}_{(L_i,l_j)\in\mathcal{G}_{\rm ARP}}$, $\{\mathrm{I}_{\rm wdt,*}(L_i,l_j)\}_{(L_i,l_j)\in\mathcal{G}_{\rm ARP}}$, $\{\mathrm{I}_{\rm wds,*}(L_i,l_j)\}_{(L_i,l_j)\in\mathcal{G}_{\rm ARP}}$, and $\{\mathrm{I}_{\rm rq,*}(L_i,l_j)\}_{(L_i,l_j)\in\mathcal{G}_{\rm ARP}}$, respectively. (e) and (f) display boxplots of the combined indices $\{\mathrm{I}_{\rm uq}(L_i,l_j)\}_{(L_i,l_j)\in\mathcal{G}_{\rm ARP}}$ and $\{\mathrm{I}_{\rm rq}(L_i,l_j)\}_{(L_i,l_j)\in\mathcal{G}_{\rm ARP}}$, respectively. The dashed lines mark the optimal index values. (g) and (h) compare maps of $\mathrm{I}_{\rm uq,U}$ and $\mathrm{I}_{\rm wdt,U}$ with $A=100$ and $300$, respectively.}
    \label{Fig:ChooseA_Supplement}
\end{figure}

This subsection provides additional details on the selection of tuning parameter $A$ in the Slepian concentration and $P$ in the VAR model. Fig.~\ref{Fig:ChooseA_Supplement}\hyperref[Fig:subfig:ChooseA_Iuq]{(a)}--\hyperref[Fig:subfig:ChooseA_Irq]{(d)} demonstrate the performance of SGs, constructed for the initial data block of length $\tau_0=365\times 8$, using various values of $A$. Across all indices, both variables display similar scales and patterns. In Fig.~\ref{Fig:ChooseA_Supplement}\hyperref[Fig:subfig:ChooseA_Iuq]{(a)} and \hyperref[Fig:subfig:ChooseA_Irq]{(d)}, $\mathrm{I}_{\rm uq,*}$ and $\mathrm{I}_{\rm rq,*}$ approach 1 and 0, respectively, as $A$ increases, stabilizing after $A\geq 300$. Comparing the maps of $\mathrm{I}_{\rm uq,U}$ with $A=100$ and $300$ in Fig.~\ref{Fig:ChooseA_Supplement}\hyperref[Fig:subfig:Iuq_A100_A300]{(g)}, we observe a significant improvement in performance at mountainous and coastal regions. These regions, as shown in Fig.~1(b), have lower standard deviations, making them more challenging to model when $A$ is small. In Fig.~\ref{Fig:ChooseA_Supplement}\hyperref[Fig:subfig:ChooseA_Iwdt]{(b)} and \hyperref[Fig:subfig:ChooseA_Iwds]{(c)}, increasing $A$ has much impact on $\mathrm{I}_{\rm wdt,*}$ and $\mathrm{I}_{\rm wds,*}$, although their ranges expand. Fig.~\ref{Fig:ChooseA_Supplement}\hyperref[Fig:subfig:Iwdt_A100_A300]{(h)} compares maps of $\mathrm{I}_{\rm wds,*}$ with $A=100$ and $300$, revealing that performance in the central region worsens as $A$ increases. As shown in Fig.~1(f), the heavy-tail characteristic in this region is relatively less pronounced compared to other regions. Applying too many coefficients with the Tukey h transformation may introduce additional error.

\begin{figure}[!t]
\centering
\subfigure[$\mathbf{P}_{U,p}$]{
\label{Fig:subfig:ChooseP_U}
\includegraphics[scale=0.65]{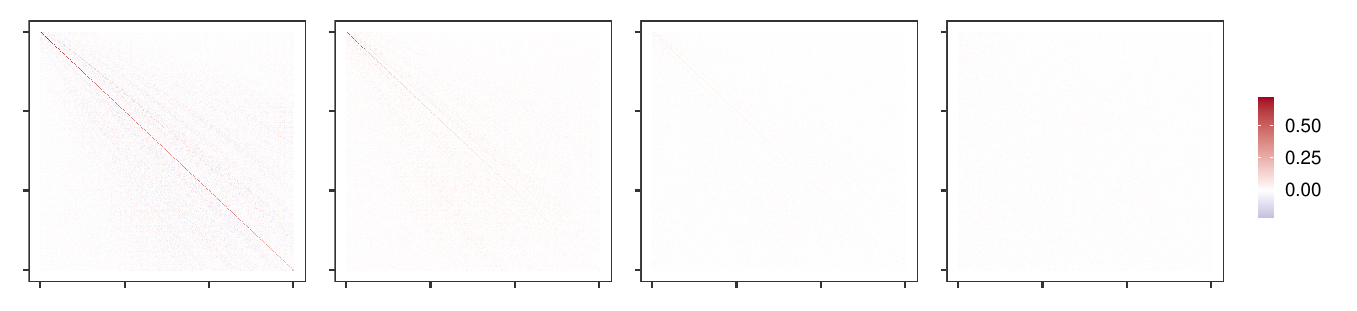}}
\subfigure[$\mathbf{P}_{V,p}$]{
\label{Fig:subfig:ChooseP_V}
\includegraphics[scale=0.65]{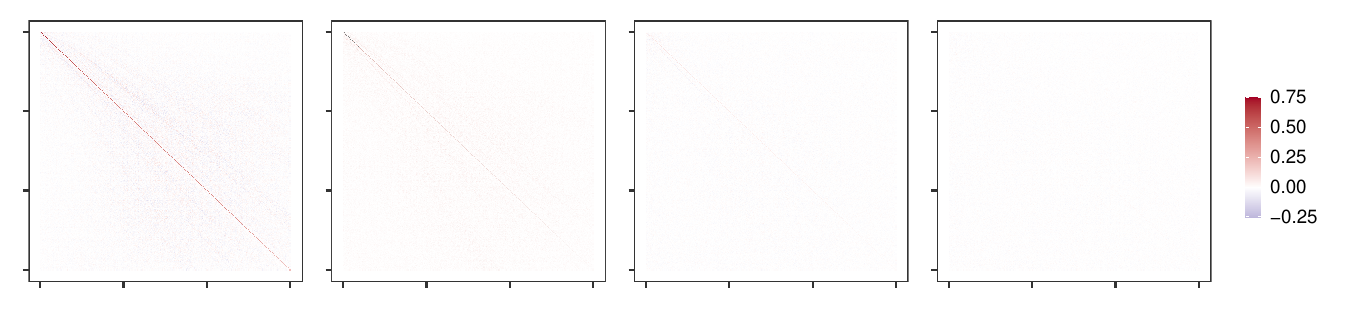}}\\
\subfigure[$\mathbf{P}_{UV,p}$]{
\label{Fig:subfig:ChooseP_UV}
\includegraphics[scale=0.65]{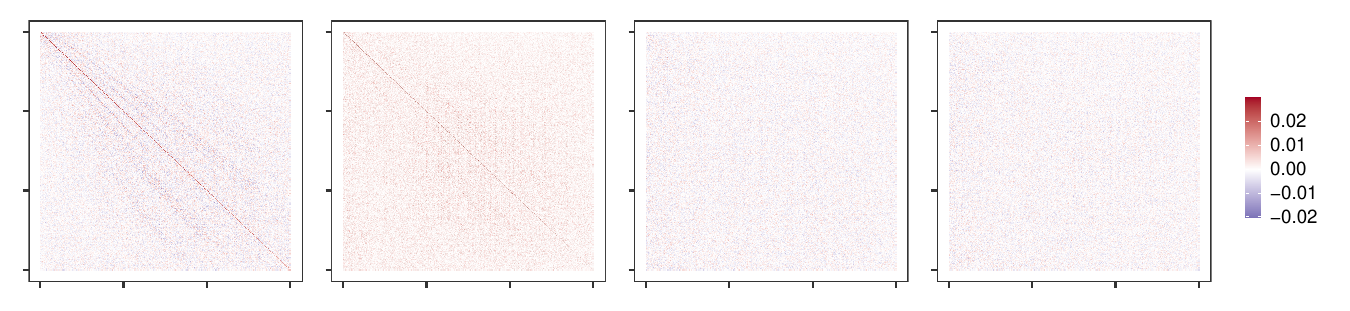}}\\
\subfigure[$\mathbf{P}_{VU,p}$]{
\label{Fig:subfig:ChooseP_VU}
\includegraphics[scale=0.65]{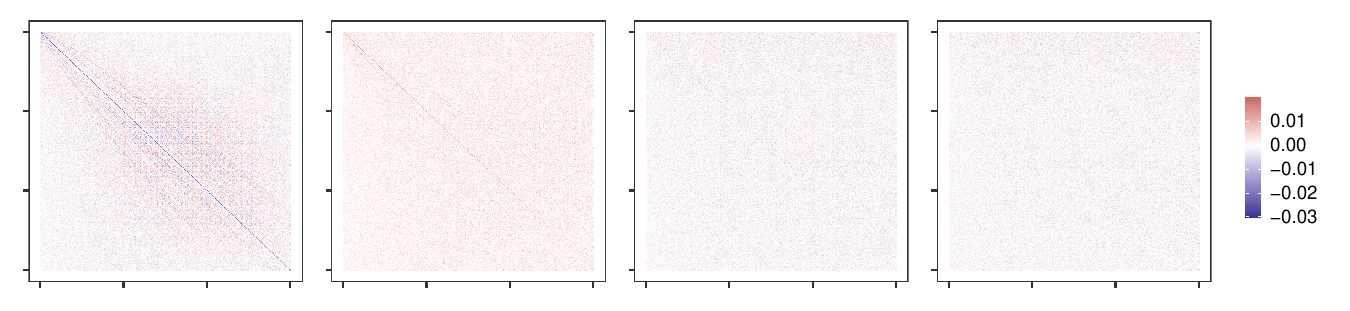}}
\caption{Selection of $P$ in the VAR model. (a)--(d) display the matrices $\mathbf{P}_{U,p}$, $\mathbf{P}_{V,p}$, $\mathbf{P}_{UV,p}$, and $\mathbf{P}_{VU,p}$, respectively, for $p=1,\ldots,4$, moving from left to right.}
\label{Fig:ChooseP_Supplement}
\end{figure}

For any time series $x_1,\ldots,x_{\tau_0}$, the $p$th order partial autocorrelation (PAC) measures the conditional correlation between $x_t$ and $x_{t-p}$, given $x_{t-1},\ldots,x_{t-p+1}$. We select $P$ as the order of the autoregressive model if a clear ``cutoff" is observed after the $P$th order of PAC. Fig.~\ref{Fig:ChooseP_Supplement} illustrates the PAC matrices of orders  $p=1,\ldots,4$. For $p=1$ and $2$, all matrices exhibit similar patterns. The values in $\mathbf{P}_{UV,p}$ and $\mathbf{P}_{VU,p}$ considerably smaller than those in $\mathbf{P}_{U,p}$ and $\mathbf{P}_{V,p}$, indicating that the temporal dependence across variables is not as strong as that within each variable. By observing the changes in patterns as $p$ increases, we choose $P=2$.

\vspace{170pt}

\subsection{Supplement to Scenario 1}
\label{sec:subsec:Supplement_to_Case1}
\begin{figure}[!h]
    \centering
    \includegraphics[scale=0.65]{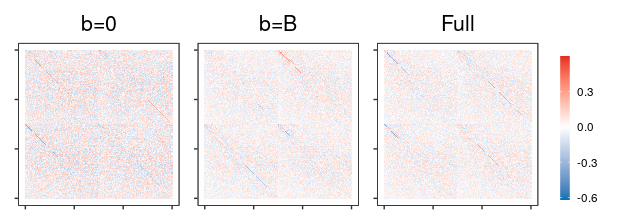}
    \caption{Estimates and updates of the matrix $\Phi_2$ in the VAR($2$) model. Rescaled cumulative estimates of $\Phi_2$ after the first block and the final block, along with its rescaled estimates obtained from the full ten-year data are illustrated. All elements have been rescaled using the function $f(x)=\mathrm{sign}(x)\sqrt{|x|}$ for clearer visualization.}
    \label{Fig:Converge_VAR2_Case1_Supplement}
\end{figure}

\begin{figure}[!h]
    \centering
    \subfigure[$\mathrm{I}_{\rm uq,V}$]{
    \label{Fig:subfig:Iuqv_Online_Case1}
    \includegraphics[scale=0.43]{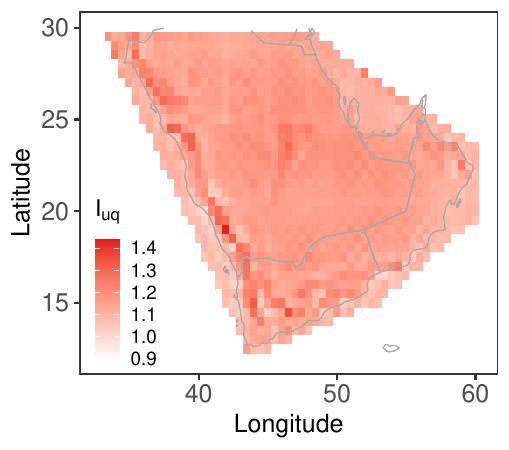}}
    \subfigure[$\mathrm{I}_{\rm wdt,V}$]{
    \label{Fig:subfig:Iwdtv_Online_Case1}
    \includegraphics[scale=0.43]{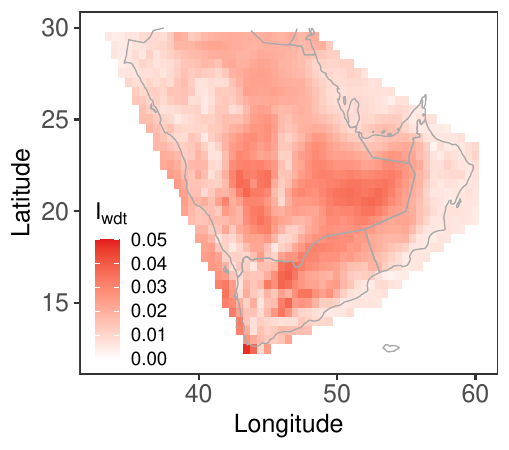}}
    \subfigure[$\mathrm{I}_{\rm wds,V}$]{
    \label{Fig:subfig:Iwdsv_Online_Case1}
    \includegraphics[scale=0.43]{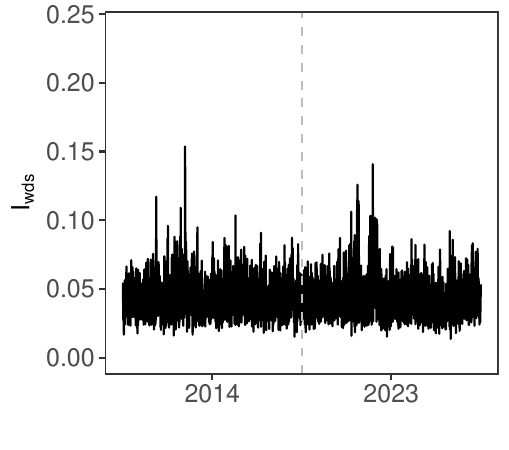}}
    \subfigure[$\mathrm{I}_{\rm rq,V}$]{
    \label{Fig:subfig:Irqv_Online_Case1}
    \includegraphics[scale=0.43]{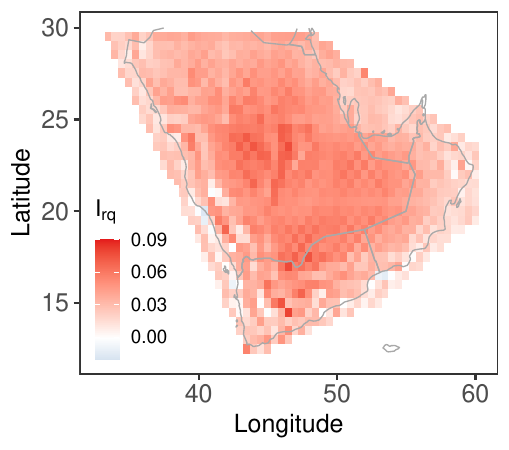}}
    \caption{Performance of the OSG constructed for the first scenario. Various metrics are presented to assess the V-component emulations generated by the OSG. }
    \label{Fig:Assessment_Case1_Supplement}
\end{figure}

As a supplement to Fig.~5,  Fig.~\ref{Fig:Converge_VAR2_Case1_Supplement} presents the estimates and updates of the matrix $\Phi_2$, which exhibit similar characteristics to those observed for $\Phi_1$. Table~\ref{tab:RFD_Case1} numerically demonstrate the convergence of cumulative estimates with increasing $b$, using relative Frobenius distances (RFDs). Fig.~\ref{Fig:Assessment_Case1_Supplement} displays the performance of the V-component emulations generated by the OSG for the first scenario. 

\begin{table}[!h]
\centering
\caption{RFDs between cumulative estimates of matrices in the VAR($2$) model at increasing $b$ values and the estimates based on the full ten-year data.}
\vspace{5pt}
{
\begin{tabular}{l|cccccccccc}
\toprule
\hline
$b$ & $0$ & $1$ & $1$ & $3$ & $4$ & $5$ & $6$ & $7$ & $8$ & $9$\\ 
\midrule
{$\Phi_1$} & $0.563$ & $0.377$ & $0.286$ & $0.232$ & $0.188$ & $0.148$ & $0.115$ & $0.093$ & $0.063$ & $0.019$\\ \hline
{$\Phi_2$} & $1.434$ & $0.958$ & $0.726$ & $0.587$ & $0.474$ & $0.373$ & $0.291$ & $0.236$ & $0.161$ & $0.048$ \\ \hline
{$\mathbf{K}$} & $0.157$ & $0.116$ & $0.074$ & $0.061$ & $0.054$ & $0.046$ & $0.033$ & $0.022$ & $0.016$ & $0.005$ \\ 
\hline
\bottomrule	
\end{tabular}}
\label{tab:RFD_Case1}
\end{table}

\subsection{Supplement to Scenario 2}
\begin{figure}[!h]
    \centering
    \subfigure[$\hat{\omega}_{V,\alpha}^{[0]}$, $\hat{\omega}_{V,\alpha}^{[B]}$, $\hat{h}_{V,\alpha}^{[0]}$, and $\hat{h}_{V,\alpha}^{[B]}$]{
    \label{Fig:subfig:OmegaH_V_alpha_Case2}
    \includegraphics[scale=0.7]{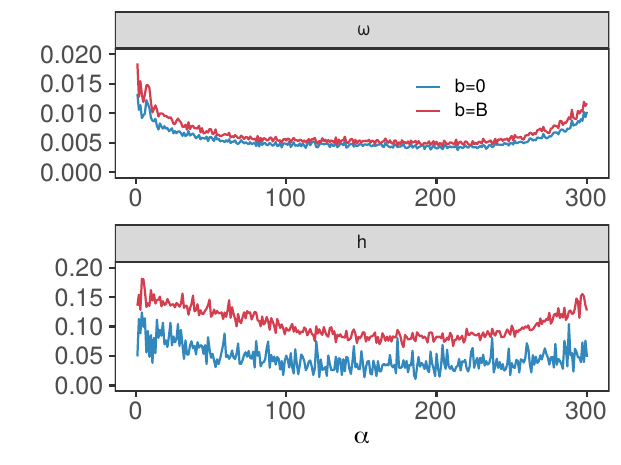}}
    \subfigure[$\hat{\omega}_{*,151}^{[b]}$ and $\hat{h}_{*,151}^{[b]}$]{
    \label{Fig:subfig:OmegaH_151_b_Case2}
    \includegraphics[scale=0.7]{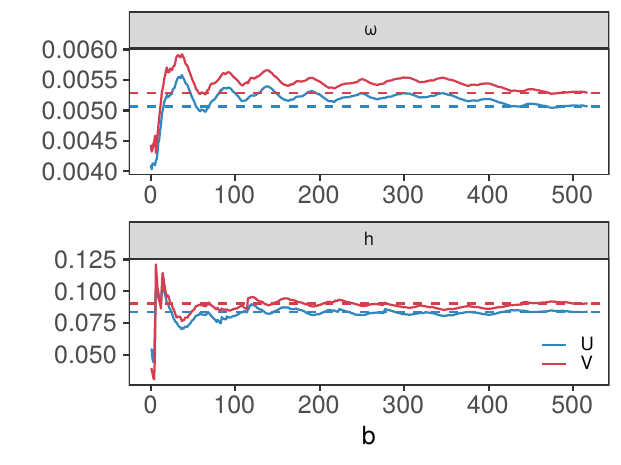}}
    \caption{Estimates and updates of parameters $\omega_{*,\alpha}$ and $h_{*,\alpha}$ in the Tukey h transformation in the second scenario. (a) compares the cumulative estimates of $\{\omega_{v,\alpha}\}_{\alpha\in\mathcal{A}}$ and $\{h_{v,\alpha}\}_{\alpha\in\mathcal{A}}$, respectively, up to the initial and the final blocks. (b) illustrates the updates of $\omega_{*,151}^{[b]}$ and $h_{*,151}^{[b]}$, respectively, over successive blocks. The dashed lines represent the estimates derived from the complete ten years of data. }
    \label{Fig:Converge_Tukeyh_Case2_Supplement}
\end{figure}

\begin{figure}[!h]
    \centering
    \subfigure[$\hat{\Phi}_2^{[b]}$ and $\hat{\Phi}_2^{\rm full}$]{
    \label{Fig:subfig:Phi2_Case2}
    \includegraphics[scale=0.65]{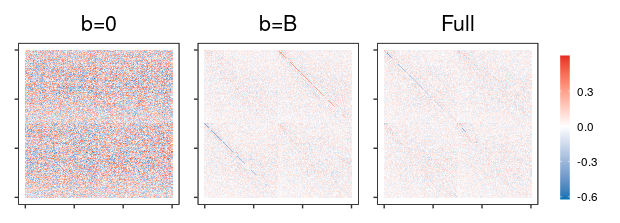}}
    \caption{Estimates and updates of the matrix $\Phi_2$ in the VAR($2$) model for the second scenario. Rescaled cumulative estimates of $\Phi_2$ up to the first block and the final block, along with its rescaled estimates obtained from the full ten-year data. All elements have been rescaled using the function $f(x)=\mathrm{sign}(x)\sqrt{|x|}$ for clearer visualization.}
    \label{Fig:Converge_VAR2_Case2_Supplement}
\end{figure}

Figs.~\ref{Fig:Converge_Tukeyh_Case2_Supplement} and \ref{Fig:Converge_VAR2_Case2_Supplement} supplement Figs.~7 and 8, respectively, by illustrating additional parameter estimates and updates in the Tukey h transformation and the VAR($2$) model.

\newpage

\begin{figure}[!h]
    \centering
    \subfigure[$\mathrm{I}_{\rm uq,V}$]{
    \label{Fig:subfig:Iuqv_Online_Case2}
    \includegraphics[scale=0.43]{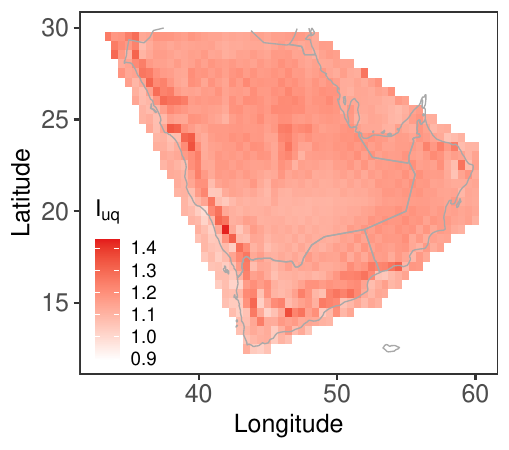}}
    \subfigure[$\mathrm{I}_{\rm wdt,V}$]{
    \label{Fig:subfig:Iwdtv_Online_Case2}
    \includegraphics[scale=0.43]{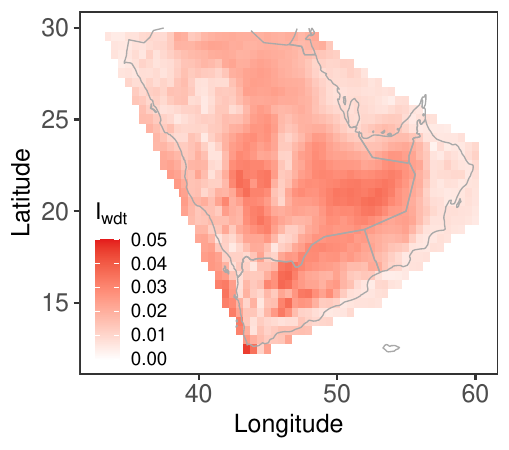}}
    \subfigure[$\mathrm{I}_{\rm wds,V}$]{
    \label{Fig:subfig:Iwdsv_Online_Case2}
    \includegraphics[scale=0.43]{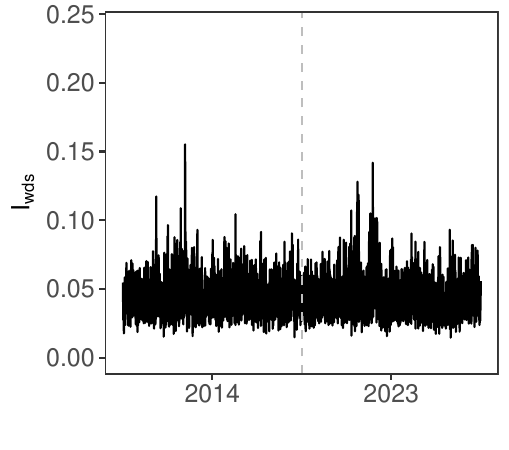}}
    \subfigure[$\mathrm{I}_{\rm rq,V}$]{
    \label{Fig:subfig:Irqv_Online_Case2}
    \includegraphics[scale=0.43]{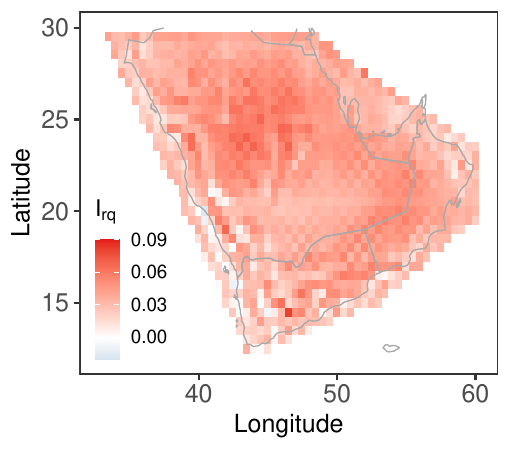}}\\
    \subfigure[$\mathrm{I}_{\rm uq,V}$]{
    \label{Fig:subfig:Boxplot_Iuqv_Case2}
    \includegraphics[scale=0.42]{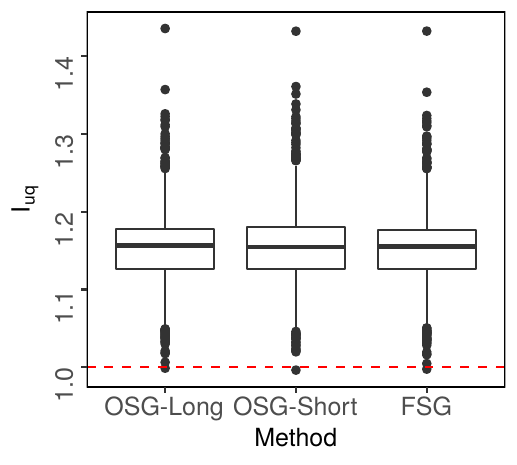}}
    \subfigure[$\mathrm{I}_{\rm wdt,V}$]{
    \label{Fig:subfig:Boxplot_Iwdtv_Case2}
    \includegraphics[scale=0.42]{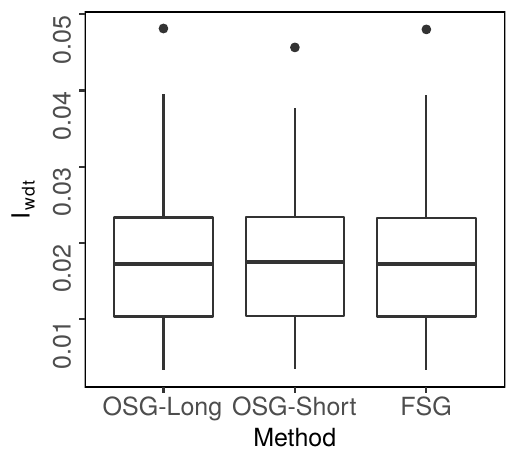}}
    \subfigure[$\mathrm{I}_{\rm wds,V}$]{
    \label{Fig:subfig:Boxplot_Iwdsv_Case2}
    \includegraphics[scale=0.42]{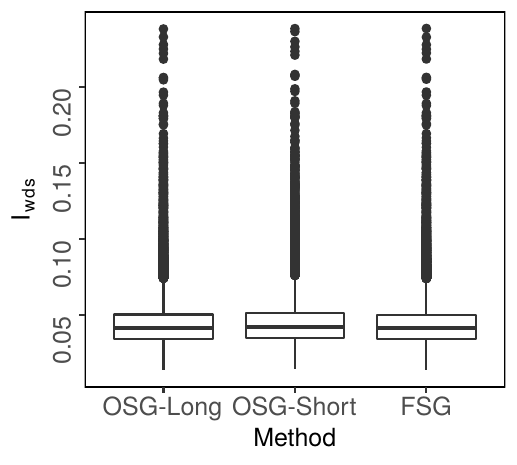}}
    \subfigure[$\mathrm{I}_{\rm rq,V}$]{
    \label{Fig:subfig:Boxplot_Irqv_Case2}
    \includegraphics[scale=0.42]{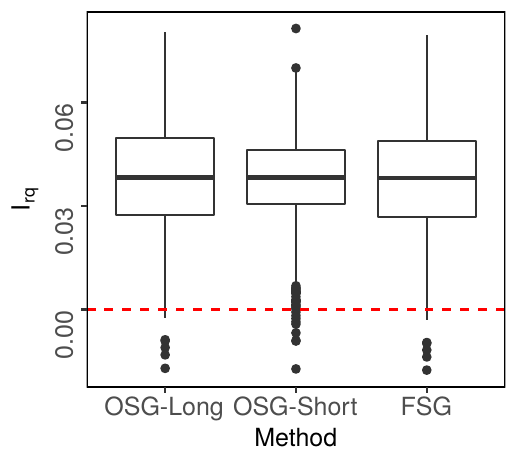}}\\
    \caption{Performance of the OSG constructed for the second scenario. (a)--(d) present various metrics used to assess the V-component emulations. (e)--(f) compare these metrics with those of the OSG-Long and the FSG. The red dashed lines in (e) and (h) mark the optimal index values.}
    \label{Fig:Assessment_Case2_Supplement}
\end{figure}

\begin{figure}[!h]
    \centering
    \subfigure[ERA5 ensemble]{
    \label{Fig:subfig:Uncertaintyv_original}
    \includegraphics[scale=0.43]{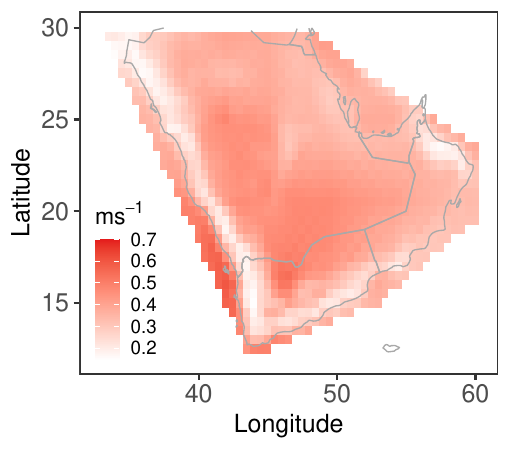}}
    \subfigure[FSG]{
    \label{Fig:subfig:Uncertaintyv_full}
    \includegraphics[scale=0.43]{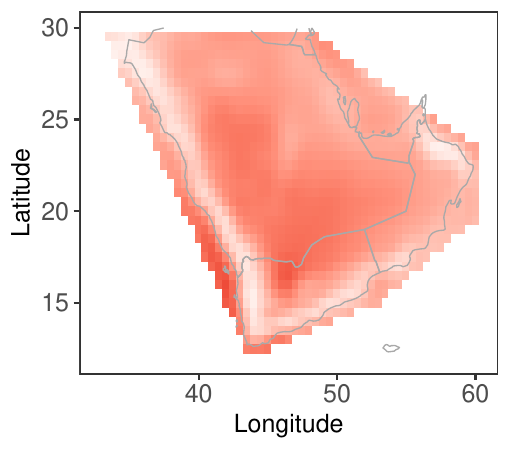}}
    \subfigure[OSG-Long]{
    \label{Fig:subfig:Uncertaintyv_Case1}
    \includegraphics[scale=0.43]{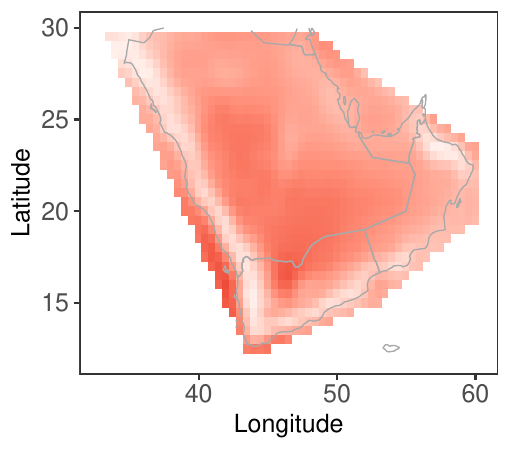}}
    \subfigure[OSG-Short]{
    \label{Fig:subfig:Uncertaintyv_Case2}
    \includegraphics[scale=0.43]{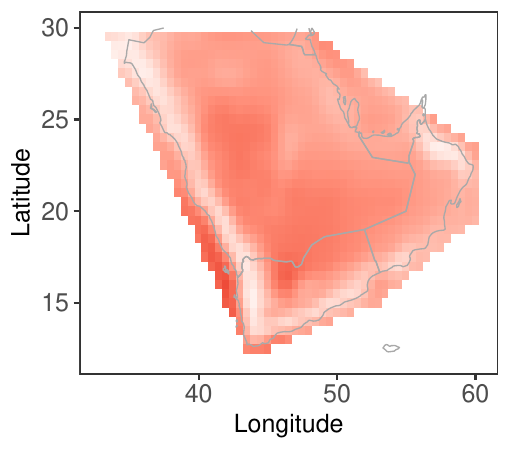}}\\
    \caption{Data unreliability over the ARP. (a) shows the map of $y_{V}^{\rm sd}(L_i,l_j)$, highlighting regions with relatively unreliable ERA5 ensembles. (b)--(d) present the maps of $\hat{y}_{V}^{\rm sd}(L_i,l_j)$.}
    \label{Fig:Uncertainty_Assessment_supplement}
\end{figure}

Figs.~\ref{Fig:Assessment_Case2_Supplement} and \ref{Fig:Uncertainty_Assessment_supplement} supplement the performance of the V-component emulations generated by the OSG-Short.

\end{document}